\documentclass[prb,twocolumn,superscriptaddress, showpacs,floatfix]{revtex4-1}  
\usepackage{hyperref}
\usepackage{amssymb}
\usepackage{amsmath}
\usepackage{float}
\usepackage[usenames]{color}
\usepackage{graphicx}
\usepackage{enumitem}
\usepackage{mathtools}
\usepackage{morefloats}
\usepackage{subfigure}

\begin{document}
\title{Second-order Green's function perturbation theory for periodic systems}
\author{Alexander A. Rusakov}
\email[Corresponding author: ]{rusakov@umich.edu}
\affiliation{Department of Chemistry, University of Michigan, Ann Arbor, Michigan 48109, USA}
\author{Dominika Zgid}
\affiliation{Department of Chemistry, University of Michigan, Ann Arbor, Michigan 48109, USA}

\begin{abstract}
Despite recent advances, systematic quantitative treatment of the electron correlation problem in extended systems remains 
a formidable task.
Systematically improvable Green's function methods capable of quantitatively describing weak and at least qualitatively strong correlations appear promising candidates for computational treatment of periodic systems.
We present a periodic implementation of temperature-dependent self-consistent 2nd-order Green's function method (GF2), where the self-energy is evaluated 
in the basis of atomic orbitals. Evaluating the real-space self-energy in atomic orbitals and solving the Dyson equation in $\mathbf{k}$-space are the key components of
a computationally feasible algorithm. We apply this technique to the 1D hydrogen lattice --- a prototypical crystalline system with a realistic Hamiltonian. 
By analyzing the behavior of the spectral functions, natural occupations, and self-energies, we claim that GF2 is able to recover metallic, band insulating, and at least qualitatively Mott regimes.
We observe that the iterative nature of GF2 is essential to the emergence of the metallic and Mott phases.

\end{abstract}
\maketitle
\section{Introduction}
In all-electron solid state calculations, density functional theory (DFT) is enormously successful.~\cite{PhysRevB.61.16440, :/content/aip/journal/jcp/125/10/10.1063/1.2347713, B812838C, PSSB:PSSB201046303, 0953-8984-24-14-145504, Dovesi_book, PhysRevB.36.891, PhysRevB.36.891, Hafner20076, JCC:JCC21057, Setyawan2010299, PhysRevLett.105.196403}
Despite its popularity, however, certain problems are persistently hard to address, such as
prediction of atomization energies and heats of formation,~\cite{doi:10.1021/cr200107z} band gaps and optical properties,~\cite{doi:10.1021/cr200107z, QUA:QUA560280846, PhysRevLett.56.2415, PhysRevB.77.115123, PhysRevLett.105.196403} as well as van der Waals interactions.~\cite{doi:10.1021/cr200107z, Kristyan1994175, :/content/aip/journal/jcp/117/24/10.1063/1.1522715}
Though new classes of exchange-correlation functionals are often able to offer superior performance, systematic improvability of DFT remains an unresolved issue.

{\em Ab initio} methods offer a systematic improvement, starting from a mean-field theory such as Hartree--Fock (HF)~\cite{Hartree, Fock, RevModPhys.23.69} 
and progressing to the hierarchies of perturbation,~\cite{doi:10.1146/annurev.pc.32.100181.002043} coupled-cluster (CC),~\cite{RevModPhys.79.291} and configuration interaction (CI)~\cite{DavidSherrill1999143} theories.
These methods, however, are prohibitively costly when applied to crystalline systems.
Additionally, being ``transplants'' of the molecular theories to periodic problems, they cannot describe the richness of the 
phases present in solids since they can diverge for metals and often prove insufficient to fully describe metal-to-insulator transitions  
due to their failure to capture a multi-reference character of the electronic states in phases such as Mott insulator.
Finally, as zero Kelvin theories, they cannot predict competing electronic phases present at finite temperatures.

Traditional {\em ab initio} methods such as HF,~\cite{Dovesi_HF, :/content/aip/journal/jcp/140/2/10.1063/1.4859257, :/content/aip/journal/jcp/143/10/10.1063/1.4930024, doi:10.1080/00268976.2010.516278} 2nd-order M\o ller--Plesset perturbation theory MP2,~\cite{:/content/aip/journal/jcp/115/21/10.1063/1.1414369, B803274M, JCC:JCC20975, C2CP23927B, :/content/aip/journal/jcp/143/10/10.1063/1.4921301, :/content/aip/journal/jcp/139/19/10.1063/1.4829898, doi:10.1021/ct400797w, :/content/aip/journal/jcp/130/18/10.1063/1.3126249, 
:/content/aip/journal/jcp/133/7/10.1063/1.3466765, doi:10.1021/jp410587b, :/content/aip/journal/jcp/140/2/10.1063/1.4859257, PhysRevB.80.085118, :/content/aip/journal/jcp/133/3/10.1063/1.3455717, :/content/aip/journal/jcp/143/10/10.1063/1.4930024, doi:10.1080/00268976.2010.516278} CCSD,~\cite{doi:10.1021/ct200263g, :/content/aip/journal/jcp/143/10/10.1063/1.4928645, :/content/aip/journal/jcp/143/10/10.1063/1.4930024, doi:10.1080/00268976.2010.516278} or the method of increments,~\cite{doi:10.1021/ct500841b, doi:10.1021/ct401040t, PhysRevB.85.045444, 0953-8984-22-27-275504, doi:10.1142/S0217979207043592, PhysRevB.46.6700, :/content/aip/journal/jcp/97/11/10.1063/1.463415} which divides the correlation energy into contributions of localized orbitals, 
recover structural and energetic properties such as cohesive (atomization) energies, lattice constants, bulk moduli, unit cell energies, and band gaps. 
However, these methods, while excellent in recovering electronic energy and some simple spectral properties like the magnitude of the band gap, are not designed to 
produce the $\mathbf k$-dependent density of states --- a property measured in angle-resolved photoelectron spectroscopy (ARPES) that remains one of the most popular experimental 
techniques for studying crystalline systems. Moreover, these methods do not yield the self-energy that allows us to distinguish band and Mott insulators and
gives additional information about the nature of correlations present in the system under consideration.

Recently, a lot of progress has been made in numerical Green's function methods, 
and currently, efficient implementations of zero-temperature random phase approximation (RPA)~\cite{doi:10.1021/ct5001268, PhysRevB.90.054115}
and GW~\cite{PhysRevB.90.075125} can be applied to solids. These methods traditionally focus on the evaluation of spectral properties such as the experimentally observed density of states.
Yet again, similarly to the {\em ab initio} methods, they cannot properly illustrate metal-to-insulator transitions and phases 
that display strong correlations.

Periodic systems are very challenging for quantum chemical/condensed matter methods, and clearly this territory is much more uncharted than the molecular realm.
Consequently, there is plenty of room for methodological and computational improvement.
In this paper, we investigate a periodic implementation of temperature-dependent 2nd-order iterative Green's function (GF2) method.~\cite{:/content/aip/journal/jcp/122/16/10.1063/1.1884965,GF2_1,GF2_2} 
To learn about the performance of GF2 for solids, we analyze a 1D infinite hydrogen lattice and 
examine the following aspects: 
{\bf (i)} the existence of multiple phases at finite temperature,
{\bf (ii)} the ability of GF2 to describe a metallic phase, 
{\bf (iii)} the description of metal-to-insulator transition and the Mott phase,
{\bf (iv)} the change in the solutions due to the iterative procedure.  
We expect that the finite temperature iterative perturbative approach is not affected by divergencies for metals in the same way as the zero-temperature 
perturbation theory such as MP2. Additionally, we expect that the iterative procedure present in GF2 is able to account for a range of correlation effects that 
cannot be described by a non-iterative method.

This paper is organized as follows. In Sec.~\ref{iterations}, we discuss the advantages of iterative Green's function methods. In Sec.~\ref{locality}, we outline the basic components of periodic GF2 and elucidate the local character of the real-space self-energy in the atomic basis.
In Sec.~\ref{procedure}, we elaborate on the details of the self-consistent procedure. 
The evaluation of the reciprocal space $\mathbf{k}$-dependent spectral function is explained in Sec.~\ref{spectra}. 
The results of our investigations for the 1D infinite hydrogen lattice are presented in Sec.~\ref{results}.
Finally, we form the conclusions in Sec.~\ref{conclusions}.

\section{Iterative Green's function methods}\label{iterations}

The system in thermal equilibrium can be described by a temperature-dependent one-body Green's function $\mathbf G(\omega_n)$
on the imaginary axis, where $\omega_n=i(2n+1)\pi/\beta$ are Matsubara frequencies, $\beta=1/(k_B T)$ is the inverse temperature, and $k_B$ is the Boltzmann constant. 
The exact expectation values of any one-body operator can be evaluated using the explicitly temperature-dependent density matrix $\hat{\gamma}=e^{-\beta(\hat{H}-\mu \hat{N})}/\operatorname{tr}(e^{-\beta(\hat{H}-\mu \hat{N})})$
where $\hat{H}$, $\hat{N}$ and $\mu$ are the system Hamiltonian, particle number operator, and chemical potential, respectively. 

In this paper, we discuss an iterative perturbative 2nd-order approximation, GF2,~\cite{:/content/aip/journal/jcp/122/16/10.1063/1.1884965, GF2_1, GF2_2} yielding an approximate
temperature-dependent density matrix from the 2nd-order Green's function.
In such method, the perturbation is not only a function of interactions, but also temperature. Therefore, GF2 is exact in the limit of infinite temperature as well as vanishing perturbation.
Recently, numerous papers appeared that investigate different variants of temperature-dependent perturbation theories,~\cite{:/content/aip/journal/jcp/143/10/10.1063/1.4930024, :/content/aip/journal/jcp/140/2/10.1063/1.4859257} since the inclusion of temperature, while not very relevant
in molecules, is crucial in materials studies.

The iterative temperature-dependent Green's function methods such as GF2 are conserving approximations,~\cite{PhysRev.124.287,:/content/aip/journal/jcp/130/11/10.1063/1.3089567} and
the observables obtained from the Green's function agree with the macroscopic conservation laws, \textit{e.g.}, conservation of particle number, momentum, angular momentum, and energy.
In such approximations, the self-energy can be expressed as a functional derivative of a certain functional with respect to the Green's function. This is only attained if the Green's function method is self-consistent.~\cite{:/content/aip/journal/jcp/122/16/10.1063/1.1884965}
If the Green's function method is not self-consistent, the particle number can be incorrect and the virial theorem does not need to be preserved. 
Most importantly, in the conserving approximations at self-consistency different methods for calculating the total energy via the Green's function give the same result.~\cite{PhysRevA.73.012511}

Another advantage of self-consistency is due to effective inclusion of higher-order diagrams, which are not explicitly present in the 2nd-order self-energy approximation, via iterative renormalization of the free propagator lines. 
In Sec.~\ref{results}, we discuss how the iterations (and implicit inclusion of higher-order diagrams) influence the phase diagram and convergence of GF2 for metallic systems.

\section{Evaluating a Green's function and self-energy in the atomic basis}\label{locality}
In this section, we discuss the strategy of building a self-consistent 2nd-order Green's function (GF2) for infinite crystalline species. To this end, we aim to simultaneously take advantage of the following useful representations of the Green's function and self-energy. 
First, in a fashion similar to Refs.~\onlinecite{GF2_1, GF2_2}, we are exploiting a convenient denominator-free form of the 2nd-order self-energy in the imaginary-time domain. 
Then, since the 2nd-order self-energy expression can be evaluated using atomic orbitals, we take advantage of the local character of the self-energy in this basis.
Third, to facilitate an efficient evaluation of the Green's function, we construct it and solve the Dyson equation in $\mathbf{k}$-space.

\subsection{The Green's function and self-energy for periodic systems: Dyson equation in the frequency domain}

For an infinite crystalline system, the evaluation of various quantities such as overlap, density and Fock matrices or Green's functions 
is made feasible by exploiting translational symmetry and imposing periodic boundary conditions (PBC).
If $i, j$ denote the indices of the functions in the atomic orbital (AO) basis, vectors indices $\mathbf{g, h}$ enumerate the real-space unit cells, and $\mathbf{0}$ is a chosen (reference) cell, then periodicity of an arbitrary
real-space matrix quantity $\mathbf{A}$ implies $A_{ij}^{\mathbf{gh}} = A_{ij}^{\mathbf{0,h-g}}$, that is: matrices resulting from a shift by an arbitrary lattice translation vector are identical.
From a computational standpoint, it means that the row of matrices $\mathbf{A}^{\mathbf{0g}}$, where $\mathbf{g}$ runs over every crystalline unit cell, contains the entire information about $\mathbf{A}$.

In the frequency domain, the general expression for the Green's function 
reads
\begin{equation}\label{GF_general}
\mathbf{G}(\omega) = \left[(\omega+\mu)\mathbf{S} - \mathbf{F} - \mathbf{\Sigma}(\omega)\right]^{-1},
\end{equation}
where $\mathbf{S}$ and $\mathbf{F}$, and $\mathbf{\Sigma}(\omega)$ are, respectively, overlap and Fock matrices, and the self-energy in the AO basis.
In this paper, we use a discrete set of Matsubara (imaginary) frequencies $\omega_n$.
Eq.~\ref{GF_general} can be written as the Dyson equation
\begin{equation}\label{Dyson_omega}
\mathbf{G}^{-1}(\omega) = \mathbf{G}_{0}^{-1}(\omega) - \mathbf{\Sigma}(\omega),
\end{equation}
where $\mathbf{G}_{0}(\omega) = \left[(\omega+\mu)\mathbf{S} - \mathbf{F}\right]^{-1}$ serves as an ``unperturbed'' 0th-order Green's function.

For a crystalline system, Eq.~\ref{Dyson_omega} turns into the following expression which facilitates the evaluation of $\left[\mathbf{G}^{-1}(\omega)\right]^{\mathbf{0g}}$ in the AO basis in the real space:
\begin{equation}\label{Dyson_omega_PBC}
\left[\mathbf{G}^{-1}(\omega)\right]^{\mathbf{0g}} = (\omega+\mu)\mathbf{S}^{\mathbf{0g}} - \mathbf{F}^{\mathbf{0g}} - \mathbf{\Sigma}^{\mathbf{0g}}(\omega).
\end{equation}
Retrieving $\left[\mathbf{G}(\omega)\right]^{\mathbf{0g}}$, however, is not attainable by the brute force inversion of the above equation, as each block would result from the inversion of the entire matrix. 
Similar problem arises in periodic HF or DFT calculations when the Fock matrix is diagonalized. Due to translational symmetry of the Green's function $\mathbf{G}^{\mathbf{0g}}(\omega) = \mathbf{G}^{\mathbf{h,g+h}}(\omega)$ though, the real-space inversion can be traditionally circumvented via a Fourier transform to 
$\mathbf{k}$-space. 
This is equivalent to symmetrization of the AO basis (see Ref.~\onlinecite{PhysRevB.61.16440} for details). In $\mathbf{k}$-space, 
the translationally invariant matrices appear in a convenient block-diagonal form:
\begin{equation}\label{FRK}
\mathbf{A}^{\mathbf{k}} = \sum_\mathbf{g} \mathbf{A}^{\mathbf{0g}} exp(i\mathbf{k} \cdot \mathbf{g}).
\end{equation}
Then solving the Dyson equation breaks down into a series of independent 
$N\times N$ matrix inversions at each $\mathbf{k}$-point sampling the first Brillouin zone, where $N$ is the number of the AO basis functions in the unit cell:
\begin{equation}\label{GFk}
\mathbf{G}^{\mathbf{k}}(\omega) = \left[(\omega+\mu)\mathbf{S}^{\mathbf{k}} - \mathbf{F}^{\mathbf{k}} - \mathbf{\Sigma}^{\mathbf{k}}(\omega)\right]^{-1}.
\end{equation}
The real-space Green's function can be then restored by inverse Fourier transform:
\begin{equation}\label{FKR}
\mathbf{G}^{\mathbf{0g}}(\omega) = \frac{1}{V_{BZ}}\sum_\mathbf{k} \mathbf{G}^{\mathbf{k}}(\omega) \operatorname{exp}(-i\mathbf{k} \cdot \mathbf{g}),
\end{equation}
where $V_{BZ}$ is the volume of the first Brillouin zone.

\subsection{Employing the real-space self-energy locality argument}\label{loc_argument}
A practical way to obtain $\mathbf{\Sigma}^{\mathbf{0g}}(\omega)$ necessary to solve Eq.~\ref{Dyson_omega_PBC} involves two steps:
$\mathbf{\Sigma}^{\mathbf{0g}}(\tau)$ is produced in the imaginary-time domain and then Fourier transformed from $\tau$ to $\omega$ domain.

In the imaginary-time domain, the 2nd-order self-energy can be compactly written as a contraction of the Green's function and two-electron integrals (for simplicity, we discuss a closed-shell case which can be generalized to open-shell systems):~\cite{GF2_1}
\begin{equation}\label{GF2_mol}
\begin{split}
\Sigma_{ij}(\tau) = -\sum_{klmnpq}G_{kl}(\tau)G_{mn}(\tau)G_{pq}(-\tau)	 \times \\ \times v_{imqk}(2v_{jnpl}-v_{jlpn}),
\end{split}
\end{equation}
where $v_{ijkl}$ are the two-electron integrals in chemists' $(11|22)$ notation (we us this notation through the rest of the text).

A generalization of this expression to the periodic case requires the following real-space blocks of $\mathbf{\Sigma}(\tau)$ to be evaluated:
\begin{equation}\label{GF2_pbc_real}
\begin{split}
\Sigma_{ij}^{\mathbf{0g}}(\tau) = -\sum_{\mathbf{g_1,\ldots,g_6}}\sum_{klmnpq}G_{k~l}^{\mathbf{g_3g_6}}(\tau)G_{m~n}^{\mathbf{g_1g_4}}(\tau)G_{p~q}^{\mathbf{g_5g_2}}(-\tau) \times \\ \times v_{i~m~q~k}^{\mathbf{0g_1g_2g_3}}(2v_{j~n~p~l}^{\mathbf{gg_4g_5g_6}}-v_{j~l~p~n}^{\mathbf{gg_6g_5g_4}}).
\end{split}
\end{equation}
In this expression, the bold indices $\mathbf{0},\mathbf{g},\mathbf{g_1},\ldots,$ $\mathbf{g_6}$ are the lattice translation vectors pointing to the unit cell where the basis function is located, and $\mathbf{0}$ is the reference cell.

In Eq.~\ref{GF2_mol}, the evaluation of the formally 6-fold contractions over the basis functions is facilitated by breaking it down into several consecutive transformations. For instance, the following intermediate contractions can be utilized: $A_{imql} = \sum_{k}G_{kl}(\tau)v_{imqk}$, 
$B_{inql} = \sum_{m}G_{mn}(\tau)A_{imql}$,  $C_{inpl} = \sum_{m}G_{pq}(-\tau)B_{inql}$, and finally $\Sigma_{ij}(\tau) =  -\sum_{npl}C_{inpl}(2v_{jnpl}-v_{jlpn})$, thus bringing the computation cost to $O(N^5)$ (we omit cell indices for simplicity).
In the periodic case, we need to produce 
$\Sigma_{ij}^{\mathbf{0g}}(\tau)$ for every cell index $\mathbf{g}$ with index $i$ pointing to the $\mathbf{0}$ cell only, thus keeping the cost of the evaluation manageable. In addition, screening the two-electron integrals and discarding negligible intermediate
contractions can reduce the computation cost even further.
	
The self-energy expression (Eq.~\ref{GF2_pbc_real}) appears as a product of quantities that decay with the increase of the intercell distance. 
Such structure of $\Sigma_{ij}^{\mathbf{0g}}(\tau)$ prompts a higher decay rate of the self-energy than that of the individual $G(\tau)$ and $v$ components as the distance between the cells increases.
This means that the self-energy is a relatively local quantity despite the summations in Eq.~\ref{GF2_pbc_real} covering the entire crystal. 

To demonstrate the local nature of the self-energy, we sketch a short numerical analysis for the one-dimensional (1D) equidistant hydrogen lattice 
with one $s$-function per atom.
Let us closer examine one of the terms in Eq.~\ref{GF2_pbc_real}
\begin{equation}\label{term} 
G_{k~l}^{\mathbf{g_3g_6}}(\tau)G_{m~n}^{\mathbf{g_1g_4}}(\tau)G_{p~q}^{\mathbf{g_5g_2}}(-\tau)v_{i~m~q~k}^{\mathbf{0g_1g_2g_3}}v_{j~n~p~l}^{\mathbf{gg_4g_5g_6}}.
\end{equation}
This term can be considered in two limits: large values of the {\bf (i)} two-electron integrals or {\bf (ii)} Green's functions.

\begin{figure}[htp]
\centering
\subfigure{\includegraphics[scale=0.7]{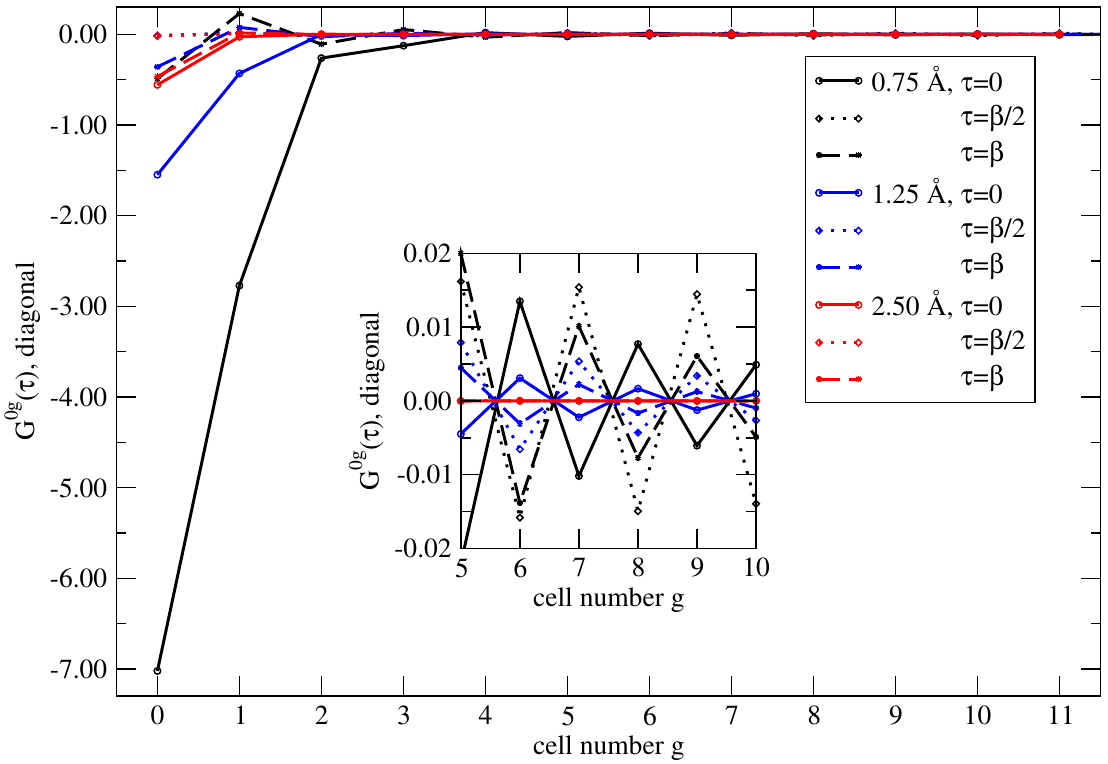}} \\
\qquad \\
\qquad \\
\subfigure{\includegraphics[scale=0.7]{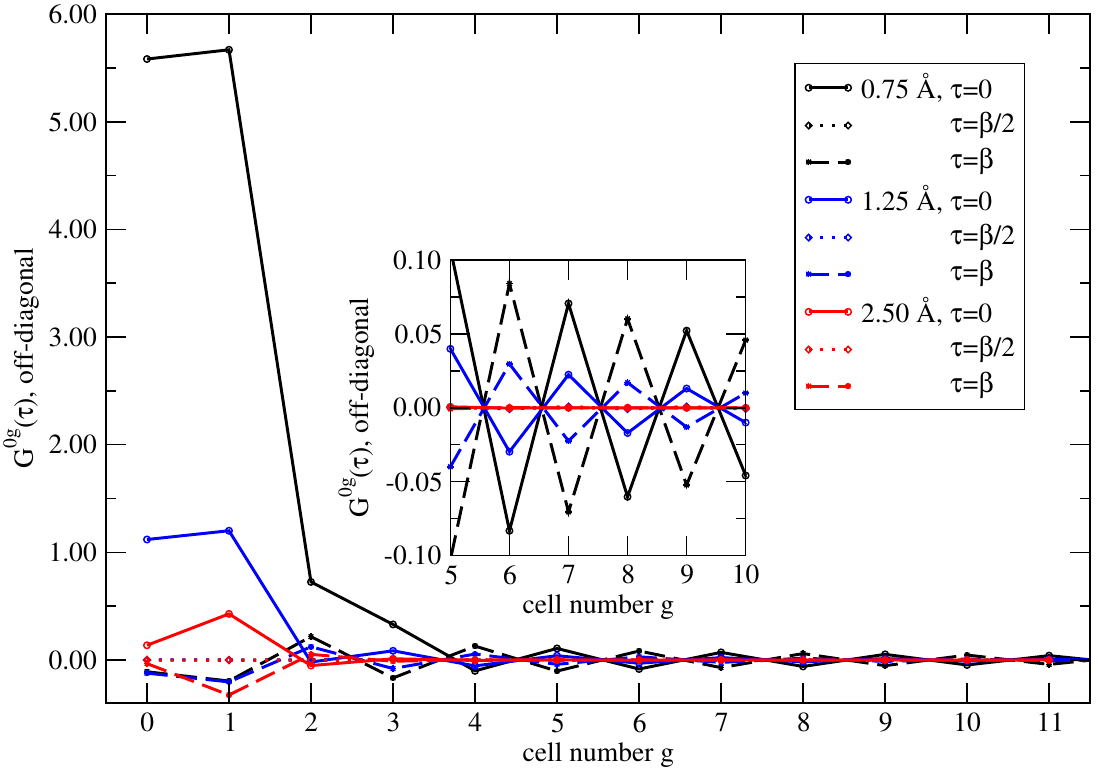}} \\
\caption{\label{GF_decay}The behavior of the Green's function matrix elements $G^{0g}_{ij}(\tau)$ (diagonal for $i=j=1$ and off-diagonal for $i=1, j=2$) for the 1D hydrogen lattices with respect to the number of cells $g$ separating the centers of basis functions $i$ and $j$. Plotted for $R=0.75$, 1.25 and 2.50~\AA~and $\tau=0$, $\beta/2$ and $\beta$ for $\beta=100$. Mini-Huzinaga basis~\cite{Huzinaga_basis} is used.}
\end{figure}

{\bf (i) Large integrals limit}.
In the chosen basis, the largest integrals are the on-site ones $v_{jjjj}^{\mathbf{gggg}}$. The corresponding absolute value of the term~(\ref{term}) is 
$|G_{ij}^{\mathbf{0g}}(\tau)G_{ij}^{\mathbf{0g}}(\tau)G_{ji}^{\mathbf{g0}}(-\tau)|v_{iiii}^{\mathbf{0000}}v_{jjjj}^{\mathbf{gggg}}$. 
Since $v_{iiii}^{\mathbf{0000}}v_{jjjj}^{\mathbf{gggg}}=v_{iiii}^{\mathbf{0000}}v_{jjjj}^{\mathbf{0000}}$, the whole expression can be written as 
\begin{equation}
|G_{ij}^{\mathbf{0g}}(\tau)G_{ij}^{\mathbf{0g}}(\tau)G_{ji}^{\mathbf{g0}}(-\tau)|v_{iiii}^{\mathbf{0000}}v_{jjjj}^{\mathbf{0000}}. 
\end{equation}
The magnitude of the above expression depends on the value of the triple Green's functions product.
In Fig.~\ref{GF_decay}, we plot a typical behavior of $G_{ij}^{\mathbf{0g}}(\tau)$ with respect to $\mathbf{g}$ for several values of the interatomic spacing parameter $R$.
Although, especially in the short bond regime ($R=0.75$~\AA), the Green's function assumes non-negligible values within 10 unit cells, 
the triple product remains of the orders of $10^{-6}$ for $G_{11}$ and $10^{-4}$ for $G_{12}$. Therefore, for the largest values of the two-electron integrals the product of the Green's functions $G_{ij}^{\mathbf{0g}}$ ensures a rapid decay of the corresponding contribution to the self-energy.

{\bf (ii) Large Green's function limit}. 
We maximize the Green's function elements in the expression (\ref{term}) by choosing the on-site elements: $G_{kk}^{\mathbf{00}}(\tau)G_{mm}^{\mathbf{00}}(\tau)G_{pp}^{\mathbf{00}}(-\tau)$.
Consequently, the expression~(\ref{term}) reduces to 
\begin{equation}
G_{kk}^{\mathbf{00}}(\tau)G_{mm}^{\mathbf{00}}(\tau)G_{pp}^{\mathbf{00}}(-\tau)v_{impk}^{\mathbf{0000}}v_{jmpk}^{\mathbf{g000}}.
\end{equation}
If we further maximize this expression by setting $v_{impk}^{\mathbf{0000}}$ to its largest value
$v_{iiii}^{\mathbf{0000}}$, we obtain
\begin{equation}
G_{ii}^{\mathbf{00}}(\tau)G_{ii}^{\mathbf{00}}(\tau)G_{ii}^{\mathbf{00}}(-\tau)v_{iiii}^{\mathbf{0000}}v_{jiii}^{\mathbf{g000}}.
\end{equation}
The integral $v_{jiii}^{\mathbf{0ggg}}$ decays fast enough to prevail over a possibly substantial product of Green's functions.
Specifically, for $R=0.75$~\AA~the largest product of the Green's functions corresponds to $|G_{ii}^{\mathbf{00}}(\tau)G_{ii}^{\mathbf{00}}(\tau)G_{ii}^{\mathbf{00}}(-\tau)|$ at $\tau = 0$, and the integral $v_{1111}^{\mathbf{0ggg}}$ drops to the order of $10^{-6}$ for $\mathbf{g}$ being 5 cells apart from the reference one. 

\begin{figure}
\centering
\subfigure{\includegraphics[scale=0.7]{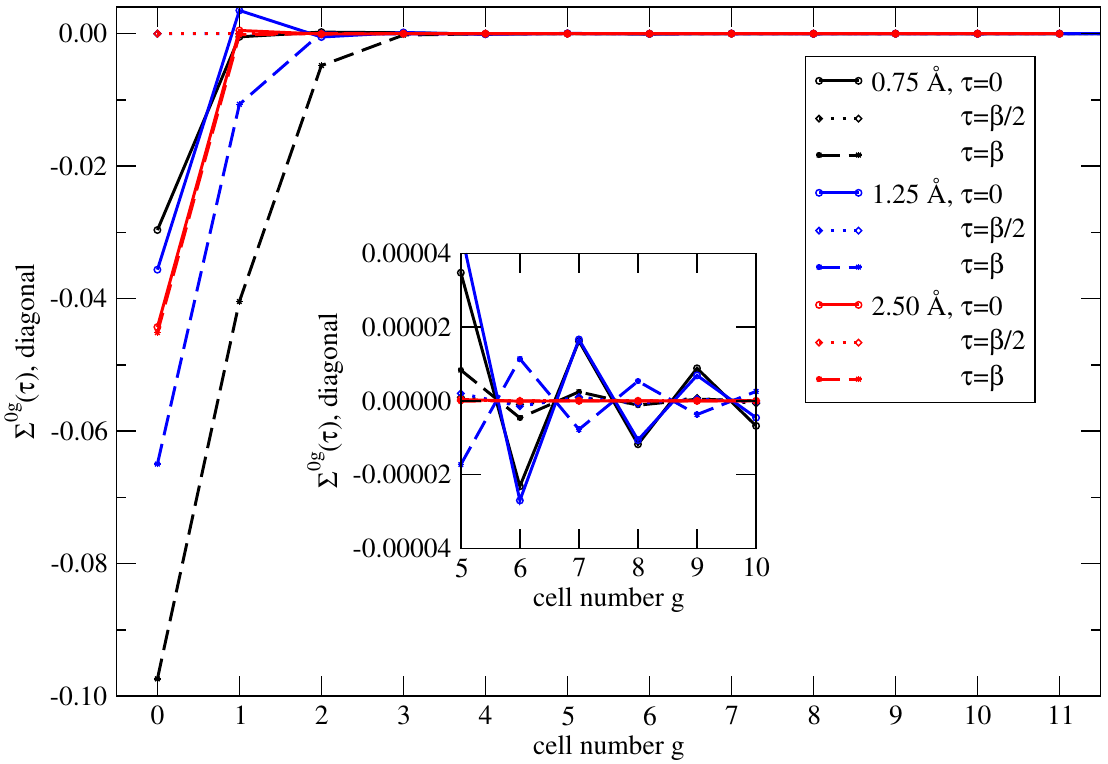}} \\
\qquad \\
\qquad \\
\subfigure{\includegraphics[scale=0.7]{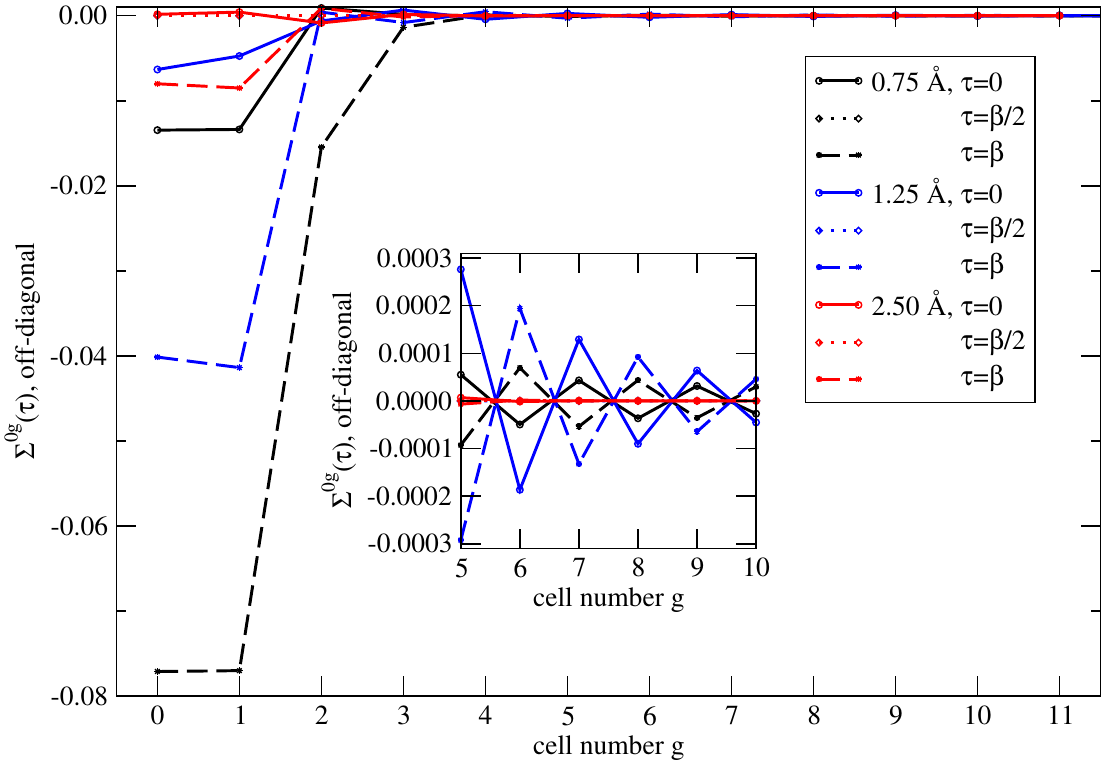}} \\
\caption{\label{SE_decay}The behavior of the self-energy matrix elements $\Sigma^{0g}_{ij}(\tau)$ (diagonal for $i=j=1$ and off-diagonal for $i=1, j=2$) for the 1D hydrogen lattices with respect to the number of cells $g$ separating the centers of basis functions $i$ and $j$. Plotted for $R=0.75$, 1.25 and 2.50~\AA~and $\tau=0$, $\beta/2$ and $\beta$; same $\beta$ and basis as in Fig.~\ref{GF_decay}. In comparison to the Green's function, the corresponding elements of the self-energy approach zero much faster with the increase of $g$.}
\end{figure}

We observe that even the largest possible elements of $G(\tau)$ or $v$ 
are contracted to facilitate a rapid decay of $\Sigma(\tau)$ with respect to the distance between cells $\mathbf{0}$ and $\mathbf{g}$.
This local behavior of the self-energy is illustrated in Fig.~\ref{SE_decay}.
In agreement with the preceding discussion, at a moderate separation between the basis functions, the self-energy elements reduce to $10^{-5}$--$10^{-4}$ in absolute value which supports the idea of the self-energy being relatively local, contrary, for example, to
the density or Fock matrices. 
Consequently, despite significant $O(N^5)$ scaling of the self-energy evaluation, the blocks of $\Sigma_{ij}^{\mathbf{0g}}(\tau)$ can be calculated for a rather moderate span of ${\mathbf{g}}$. 
The resulting $\Sigma_{ij}^{\mathbf{0g}}(\tau)$ can be finally Fourier transformed to $\Sigma_{ij}^{\mathbf{0g}}(\omega)$  to solve the Dyson equation. The transformations of the Green's function and self-energy between the imaginary-frequency and imaginary-time domains are made computationally efficient via a representation of the imaginary-time quantities 
in the basis of orthogonal (Legendre) polynomials.~\cite{PhysRevB.84.075145, Kananenka_Legendre}

\section{Self-consistent evaluation of the 2nd-order Green's function}\label{procedure}
At this point, we can assemble the components described in the previous sections 
in a self-consistent procedure for the crystalline Green's function with the 2nd-order self-energy.
The self-consistent GF2 procedure within the PBC framework, regardless of a particular implementation, has the same basic ingredients as the molecular one~\cite{GF2_1} 
and can be abstractly viewed as a double loop.
\begin{figure}
\centering
\includegraphics[scale=0.4]{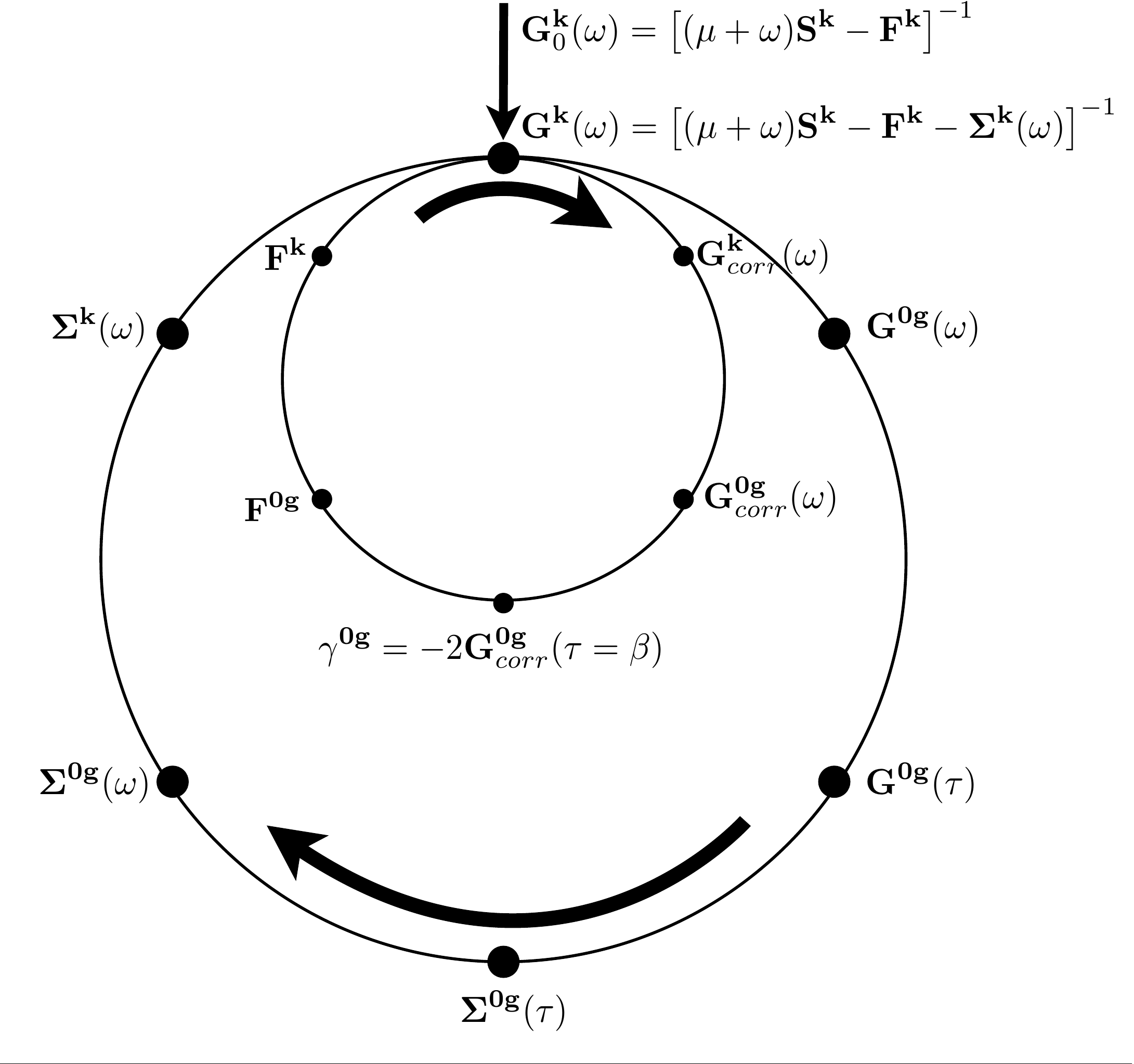}
\caption{\label{cycle}GF2 self-consistency cycle.}
\end{figure}
\begin{enumerate}[leftmargin=0cm,itemindent=.5cm,labelwidth=\itemindent,labelsep=0cm,align=left,label=(\alph*)]
\item In the greater loop, the frequency-dependent 2nd-order self-energy is evaluated from a given Green's function and two-electron integrals, 
according to Eq.~\ref{GF2_pbc_real}. 
The procedure is initiated with 
$\mathbf{G}_{0}^{\mathbf{k}}(\omega) = \left[(\omega+\mu)\mathbf{S}^{\mathbf{k}} - \mathbf{F}^{\mathbf{k}}\right]^{-1}$
where
$\mathbf{\Sigma}(\omega) = 0$ and $\mathbf{F}^{\mathbf{k}}$ typically comes from a converged HF of DFT calculation. 
Since the Green's function entering the loop is in the $\mathbf{k}$-space, an extra step 
$\mathbf{G}^{\mathbf{k}}(\omega)$ (from Eq.~\ref{GFk}) $\xrightarrow{Eq.~\ref{FKR}} \mathbf{G}^{\mathbf{0g}}$
is necessary, compared to the molecular case. 
Most generally, once $\mathbf{\Sigma}^{\mathbf{0g}}(\tau)$ is evaluated and Fourier transformed to $\mathbf{\Sigma}^{\mathbf{0g}}(\omega)$, 
it should undergo the transformation $\mathbf{\Sigma}^{\mathbf{0g}}(\omega) \xrightarrow{Eq.~\ref{FRK}} \mathbf{\Sigma}^{\mathbf{k}}(\omega)$  
to enter a new Green's function evaluation at the common point of the two loops. 

\item In the lesser loop, the frequency-independent term $\mathbf{\Sigma_{\infty}}$ is evaluated.

\end{enumerate}

The evaluation of $\mathbf{\Sigma_{\infty}}$ can be cast as constructing a Fock-type matrix $\mathbf{F} =\mathbf{h}+\mathbf{J}(\gamma)-\frac{1}{2}\mathbf{K}(\gamma) = \mathbf{h}+ \mathbf{\Sigma_{\infty}}$ from a correlated density matrix $\mathbf{\gamma} = \mathbf{\gamma}_{corr}$, where $\mathbf{h}$ is the core Hamiltonian, $\mathbf{J}(\gamma)$ and $\mathbf{K}(\gamma)$ are the Coulomb and exchange operators, respectively. 
In practical realization of the method, any quantum chemistry software capable of treating crystalline problems can be used to take the correlated density as input 
and perform a single iteration (\textit{i.e.}, without reaching the self-consistency) of the regular HF procedure to build $\mathbf{F}$ from a given $\mathbf{\gamma}$.
The correlated density matrix is evaluated as $\gamma_{corr}=-2\mathbf{G}_{corr}(\tau = \beta)$. To this end,
the  correlated Green's function $\mathbf{G}_{corr}^{\mathbf{k}}(\omega)$ in the $\mathbf{k}$-space from Eq.~\ref{GFk} 
 is transformed to the real space $\mathbf{G}_{corr}^{\mathbf{0g}}(\omega)$ via Eq.~\ref{FKR} and then to the imaginary-time domain at $\tau=\beta$. 

It is essential to assure that the resulting correlated density yields a correct number of electrons per unit cell. 
This is enforced by the chemical potential $\mu$ search: the value of $\mu$ is
adjusted in such way that the correlated density matrix $\mathbf{\gamma}^{\mathbf{0g}}_{corr} = -2\mathbf{G}^{\mathbf{0g}}_{corr}(\tau = \beta)$ 
contracted with the overlap matrix preserves
the correct number of electrons in the unit cell $N_e = \sum_{\mathbf{g}, i, j}\mathbf{\gamma}^{\mathbf{0g}}_{ij} \cdot \mathbf{S}^{\mathbf{0g}}_{ij}$.

Now, $\mathbf{F}=\mathbf{h} + \mathbf{\Sigma_{\infty}}$ can be evaluated as a single iteration of the HF procedure where a correlated density matrix is supplied to be contracted with the two-electron integrals. 
To close the lesser loop, one transforms $\mathbf{F}^{\mathbf{0g}}$ to the $\mathbf{k}$-space.
At this point, all components are ready to continue to the greater loop until convergence is reached.

The change in the electronic energy per unit cell during iterations is used as a convergence criterion. 
This energy is a sum of two terms resulting 
from the frequency-independent and frequency-dependent parts of the self-energy. 
The frequency-independent part  yields the ``one-body'' energy evaluated as
\begin{equation}\label{Sigma_inf}
E_{1b} = \frac{1}{2} \sum_{\mathbf{g}, i, j}\mathbf{\gamma}^{\mathbf{0g}}_{ij}(2\mathbf{h}^{\mathbf{0g}}_{ij}+\left[\mathbf{\Sigma}_{\mathbf \infty}\right]^{\mathbf{0g}}_{ij})
\end{equation}
using the \textit{correlated} density matrix.
The frequency-dependent part results in the ``two-body'' contribution
\begin{equation}\label{Galitskii}
E_{2b} = \frac{2}{\beta}\sum_{\mathbf{g}, i, j}\operatorname{Re}\left[\sum_{\omega} \mathbf{G}^{\mathbf{0g}}_{ij}(\omega)\mathbf{\Sigma}^{\mathbf{0g}}_{ij}(\omega) \right],
\end{equation}
where $\mathbf{G}^{\mathbf{0g}}_{ij}(\omega)$ is the \textit{correlated} Green's function.

Let us mention that using the machinery we have just presented, one can also readily access temperature-dependent MP2 energy for a periodic system. 
According to Ref.~\onlinecite{:/content/aip/journal/jcp/93/8/10.1063/1.459578}, the MP2 correlation energy per unit cell is 
\begin{equation}\label{MP2}
E_{2b, MP2} = \frac{1}{\beta}\sum_{\mathbf{g}, i, j}\operatorname{Re}\left[\sum_{\omega}\left[\mathbf{G}_0(\omega)\right]^{\mathbf{0g}}_{ij}\mathbf{\Sigma}^{\mathbf{0g}}_{ij}(\omega)\right],
\end{equation}
where $\mathbf{G}_{0}(\omega)$ results from the HF Green's function $\mathbf{G}^{\mathbf{k}}_{0}(\omega) = \left[(\omega+\mu)\mathbf{S^{k}} - \mathbf{F^{k}} \right]^{-1}$, 
and the self-energy is evaluated
in a single run of the GF2 cycle using this starting HF Green's function $\mathbf{G}_{0}(\omega)$. 
The evaluation of the MP2 energy is not a self-consistent procedure, and the total energy comes out as
$E_{HF} + E_{2b, MP2}$.

\section{Evaluation of spectral functions}\label{spectra}

Matsubara Green's functions, in contrast to the real-axis ones, are smooth and convenient to use in iterative schemes, but they
do not provide a direct access to the spectral properties of a system. 
Analytic properties of the Matsubara Green's function on the imaginary axis assure the existence and uniqueness of its analytical continuation to the real frequency axis. 
Numerical analytical continuation of the Matsubara data is an ill-posed problem,~\cite{Silver} but it is required for obtaining spectral functions from imaginary-time algorithms.~\cite{Gubernatis, RevModPhys.83.349}

In this work, we use the maximum entropy analytical continuation method as implemented in Ref.~\onlinecite{AC_ME_APLS} to obtain the spectral function.
The spectral function is defined as
\begin{equation}\label{specfunc}
A(\mathbf{k}, \omega) = -\frac{1}{\pi}\operatorname{Tr}\left[{\rm Im}\mathbf{G^k}(\omega)\mathbf{S^k}\right],
\end{equation}
where $\mathbf{G^k}(\omega)$ is the real-frequency Green's function, and $\mathbf{S^k}$ is the overlap matrix introduced to generalize the expression to a non-orthogonal basis. 
The spectral function $A(\mathbf{k}, \omega)$  is a momentum-dependent density of states and  
gives the number of states available to the electrons with energy $\omega$ and momentum $\mathbf{k}$. 
  
The success of constructing the spectral function can be verified via the ``back continuation'' to the imaginary axis by calculating the following integral:
\begin{equation}\label{back}
G^{\mathbf{k}}(\omega_n) = \int_{-\infty}^{+\infty}d\omega\frac{A(\mathbf{k}, \omega)}{\omega_n - \omega},
\end{equation} 
where $\omega_n$ stands for the $n$th Matsubara frequency.

\section{Electron correlations in one-dimensional periodic hydrogen}\label{results}

One-dimensional (1D) periodic hydrogen is the simplest crystalline system described by a realistic Hamiltonian which makes it suitable for testing the periodic GF2 method. 
Despite its simplicity, depending on the interatomic distances (corresponding to different pressures) such system displays three phases:
a metal, a band insulator, and a Mott insulator.

In this study, we employed the mini-Huzinaga basis set~\cite{Huzinaga_basis} that has one $s$-function per hydrogen atom. 
To remain within the spin-restricted formulation of GF2, we choose two hydrogen atoms per unit cell and assume 
a closed-shell one-electron reference throughout this discussion. In all Green's function evaluations we use 5000 Matsubara frequencies and the
inverse temperature $\beta = 100$ [1/a.u.] (corresponding to 3157.8 K).
The update of the frequency-independent part of the self-energy present in the GF2 method is executed via running a single iteration of HF using the  
 \textsc{gaussian}~\cite{g09} program that takes the correlated density matrix as input. 

\subsection{Convergence of the electronic energy}
The GF2 energy is calculated as $E_{1b}+E_{2b}$, where $E_{1b}$ and $E_{2b}$ are defined in Eqs.~\ref{Sigma_inf} and \ref{Galitskii}. The one-body energy (Eq.~\ref{Sigma_inf}) is evaluated after the Fock matrix is created using the correlated density matrix; the cell index $\mathbf g$ runs over the entire crystal.   
Similarly, the two-body energy, $E_{2b}$ (Eq.~\ref{Galitskii}), formally requires a summation over the infinite number of cell indices $\mathbf{g}$.
In Sec.~\ref{loc_argument}, we have illustrated that the frequency-dependent self-energy is quite local.
Therefore, it is natural to expect that a relatively moderate span of cell indices $\mathbf{g}$ is required to converge the two-body energy sum in Eq.~\ref{Galitskii}.
Since the evaluation of $\mathbf{\Sigma}(\tau)$ is the computational bottleneck of the GF2 calculation, a small range of the cell indices $\mathbf{g}$ is crucial for making our calculations computationally affordable.

To assess the effect of truncating the summation, we evaluated the energy from Eq.~\ref{Galitskii} as a function of the number of cells.
In Table~\ref{Corr_conv}, we illustrate the unit cell $E_{2b}$ behavior for three spacings of the lattice from tight (0.75 \AA) to relatively wide (2.50 \AA). 
The presented values correspond to a converged GF2 iterative
procedure with the GF2 energy convergence criterion $10^{-6}$ a.u.
\begin{table}[h!]
\centering
\begin{tabular}{c|c|c|c} 
 \hline
 & \multicolumn{3}{c}{R, \AA} \\
\#$N$ & 0.75 & 1.25 & 2.50  \\ 
 \hline\hline
 1 & $-0.205953$  &  $-0.104473$     &   $-0.127235$      \\ 
 2 & $-0.120203$  &  $-0.078355$     &   $-0.128165$      \\
 3 & $-0.034454$  &  $-0.052237$    &   $-0.129095$      \\
 4 & $-0.035047$  &  $-0.052109$     &  $-0.129024$       \\
 5 & $-0.035640$  &   $-0.051980$    &   $-0.128953$      \\
 6 & $-0.035553$  &  $-0.051918$     &   $-0.128949$      \\
 7 & $-0.035466$  &  $-0.051855$     &   $-0.128944$      \\
 8 & $-0.035478$  &  $-0.051830$     &  $-0.128944$       \\
 9 & $-0.035490$  &   $-0.051805$    &   $-0.128943$      \\
 10 & $-0.035495$  &  $-0.051793$     &   $-0.128943$      \\
 11 & $-0.035500$  &   $-0.051781$    &  $-0.128943$       \\
 \hline
\end{tabular}
\caption{The convergence of $E_{2b}$, a.u., per unit cell with respect to \#$N$ --- the number of unit cells used in Eq.~\ref{Galitskii}.}
\label{Corr_conv}
\end{table}

Even for the most tight spacing $R=0.75$~\AA,~$E_{2b}$ reaches the $10^{-3}$ a.u. plateau for 3 cells and $10^{-5}$ a.u. for 11 cells, 
while the same convergence criteria for the $E_{1b}$ term are only met for at least twice as many cells. 
Thus, while with the increase of the intercell distance the number of cells required to converge $E_{2b}$ rapidly drops, the number of cells necessary to converge $E_{1b}$ decreases quite slowly: for $R=0.75$, 1.25, and 2.50~\AA~converging  $E_{1b}$ to $10^{-5}$ a.u. requires, respectively, 25, 21, and 17 unit cells. 
Although no universal recipe can be derived from this simple case, the convergence rate of the $E_{2b}$ part of the GF2 energy with respect to the number of cells explicitly summed over in Eq.~\ref{GF2_pbc_real} 
is fast due to the local nature of the 2nd-order self-energy and can lead to computational time savings.

\subsection{1D equidistant hydrogen energy curve}
\begin{figure}
\includegraphics[scale=0.8]{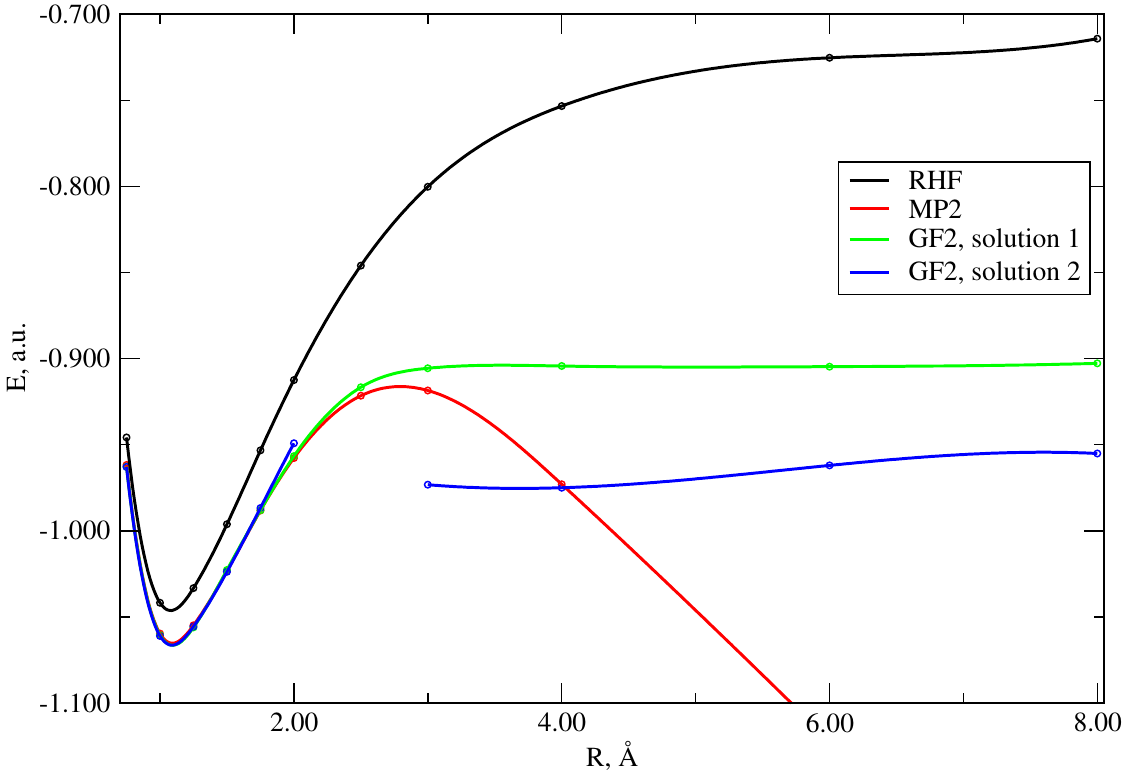}
\caption{\label{energy} Equidistant 1D hydrogen stretching curves: electronic energies per unit cell for restricted closed-shell Hartree--Fock (RHF), 2nd-order M\o ller--Plesset perturbation theory (MP2), and two GF2 solutions.}
\end{figure}

We examine an equidistant stretching of the 1D periodic hydrogen lattice. HF and MP2 energy curves in Fig.~\ref{energy} 
display features that are qualitatively similar to those previously observed in Refs.~\onlinecite{GF2_1, :/content/aip/journal/jcp/93/8/10.1063/1.459578}. 
Specifically, closed-shell HF significantly overestimates the unit cell energy at dissociation, while MP2 diverges for interatomic distances exceeding 2~\AA.

In an equidistant stretch of the 1D periodic hydrogen lattice, multiple phases with different Helmholtz free energies can be present.
A GF2 calculation, where the Green's function is evaluated on the Matsubara axis, yields the electronic part of the internal energy.
However, since GF2 is a nonlinear iterative procedure, for a given crystalline geometry it can converge to different energy solutions depending on the starting point.
The existence of multiple solutions in nonlinear iterative methods is not uncommon and is known to transpire in 
the HF~\cite{PhysRevLett.101.193001} and GW~\cite{PhysRevB.92.115125, PhysRevB.86.081102} methods.
Thus, to determine the most stable phase based on Helmholtz free energy, the electronic entropy contribution needs to be evaluated.
We will refrain from evaluating the entropic contribution in this publication and address it separately. 
Here we will focus on analyzing possible phases and phase transitions based solely on the electronic internal energy.

Although multiple solutions are possible, in this particular calculation,
we have observed convergence to two solutions.
The green curve in Fig.~\ref{energy} (``solution~1'') corresponds to the solution obtained using the HF initial guess. 
Superficially, this curve resembles the behavior of the finite species in Ref.~\onlinecite{GF2_1}. 
The blue curve in Fig.~\ref{energy} (``solution~2'') is obtained from a starting point that assumes 
no interactions between hydrogen atoms (``atomic'' initial $\mathbf{F^k}$ matrices).
In the region close to equilibrium, both GF2 solutions are practically indiscernible from MP2.
Once MP2 breaks down around past 2~\AA, the blue curve (``solution~2'') separates from the green one (``solution~1'') and stays consistently lower in energy until complete atomization. 
We observe that around 2~\AA~the blue ``solution~2'' undergoes a phase transition, and we experience difficulties converging GF2 near the geometry where it happens (this explains a blank region on the blue curve around 2.5~\AA). 

In the next sections, based on occupation numbers we will show that both GF2 solutions undergo phase transitions during the equidistant stretch.
Since the metallic, band insulating, and Mott insulating character of a solution cannot be concluded from the energy profile alone, 
we will turn to analyzing spectral functions, self-energies, and occupation numbers for ``solution~1'' and ``solution~2''.

\subsection{Self-energies and spectral functions of the 1D hydrogen lattice}
We present the self-energies and spectral functions since these behave differently for different phases 
and provide a unique characterization of a solution.

We plot the angle-resolved spectral function $A(\mathbf{k}, \omega)$ both as a 3D plot and a 2D projection on the $k$ -- $\omega$ plane using matplotlib.~\cite{Hunter:2007}
The color of the projection indicates the ``height'' of the original spectral function. 
These projections can be interpreted as remnants of the bands similar to the ones resulting from one-body methods such as HF or DFT. 
However, it is important to note that these correlated bands do not have the same meaning as 
bands in the one-electron picture since they emerge from a many-body calculation.

\begin{figure}[htp]
\centering
\subfigure{\includegraphics[scale=1.0]{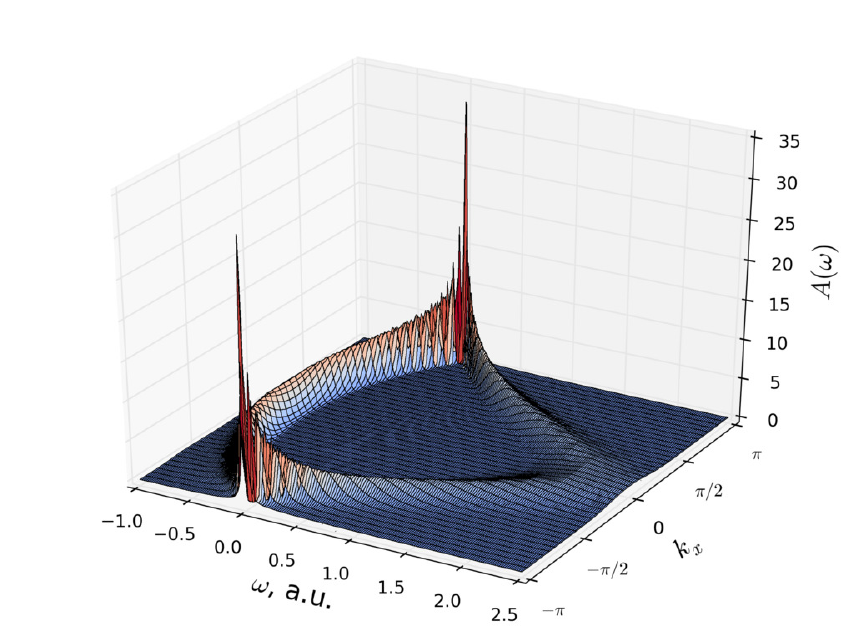}} \\
\qquad
\subfigure{\includegraphics[scale=0.9]{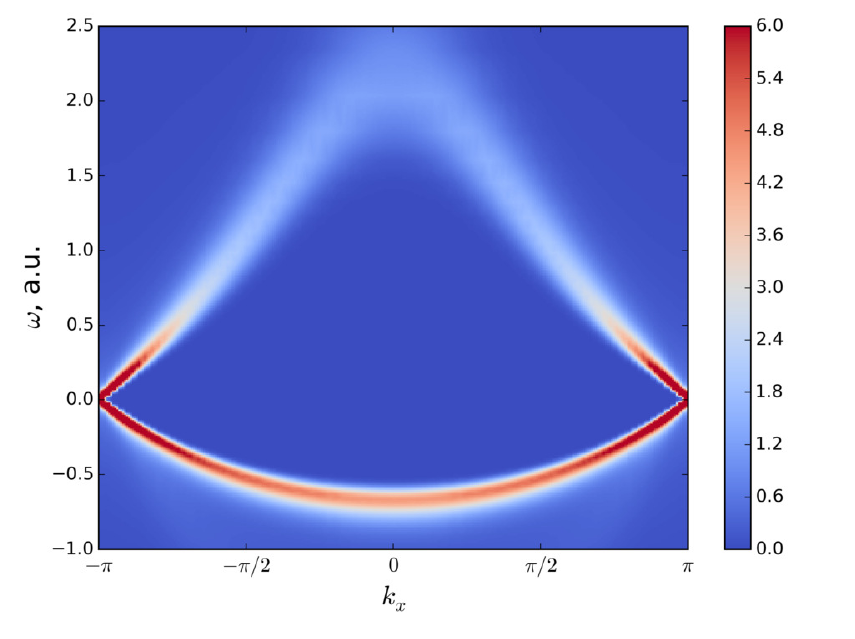}} \\
\qquad \\
\qquad \\
\subfigure{\includegraphics[scale=0.7]{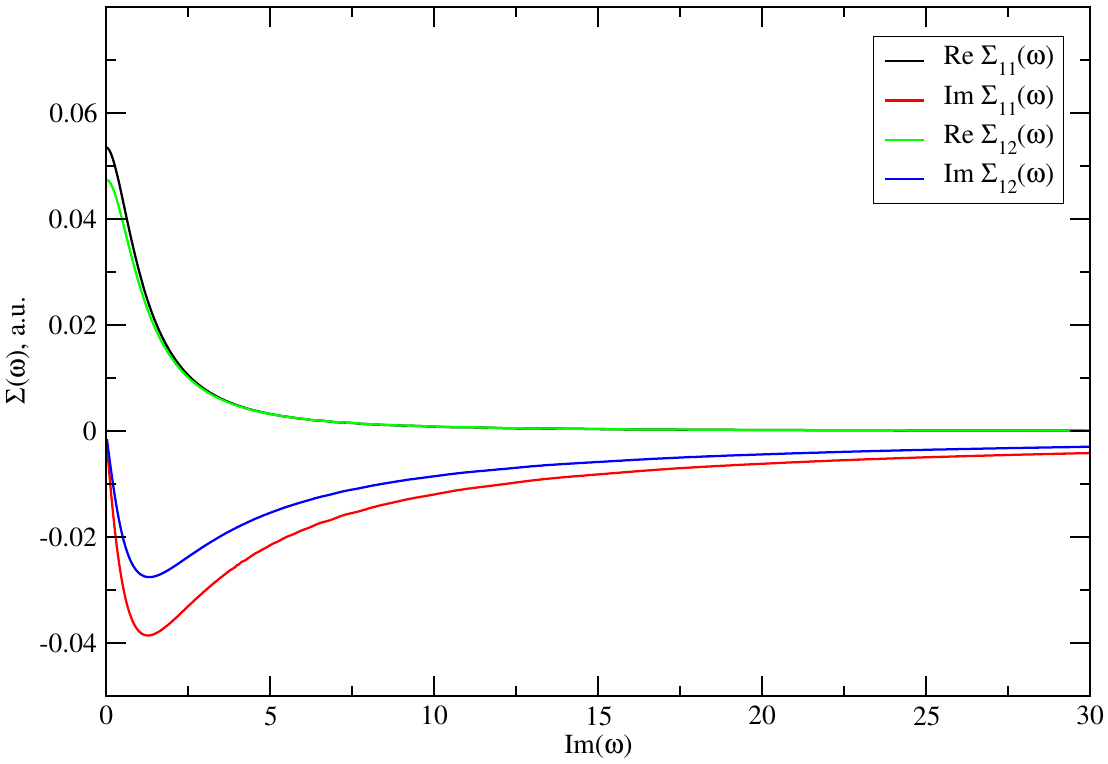}}
\caption{\label{075_1} Real-frequency spectral function and correlated bands, Matsubara self-energy for ``solution~1'' at $R=0.75$~\AA.}
\end{figure}
\begin{figure}[htp]
\centering
\subfigure{\includegraphics[scale=1.0]{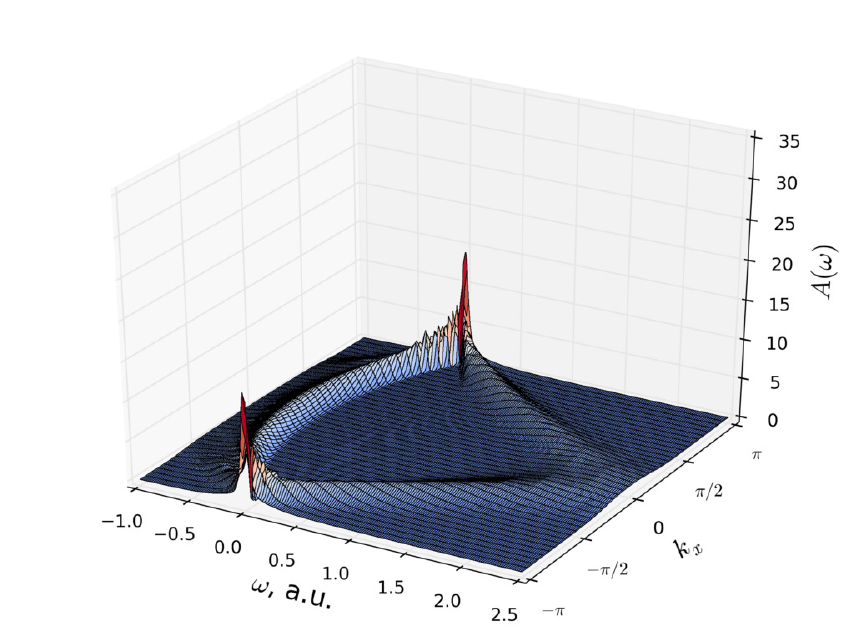}} \\
\qquad
\subfigure{\includegraphics[scale=0.9]{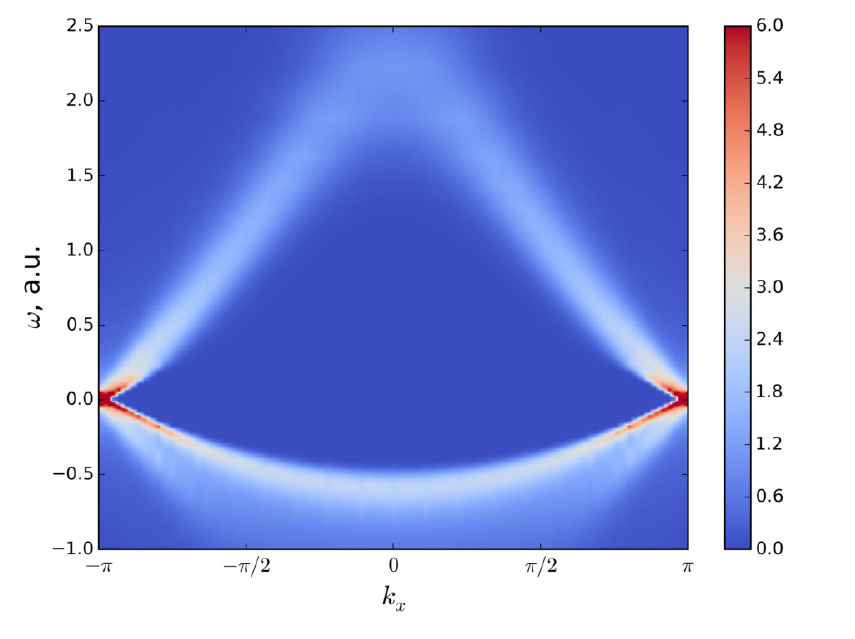}} \\
\qquad \\
\qquad \\
\subfigure{\includegraphics[scale=0.7]{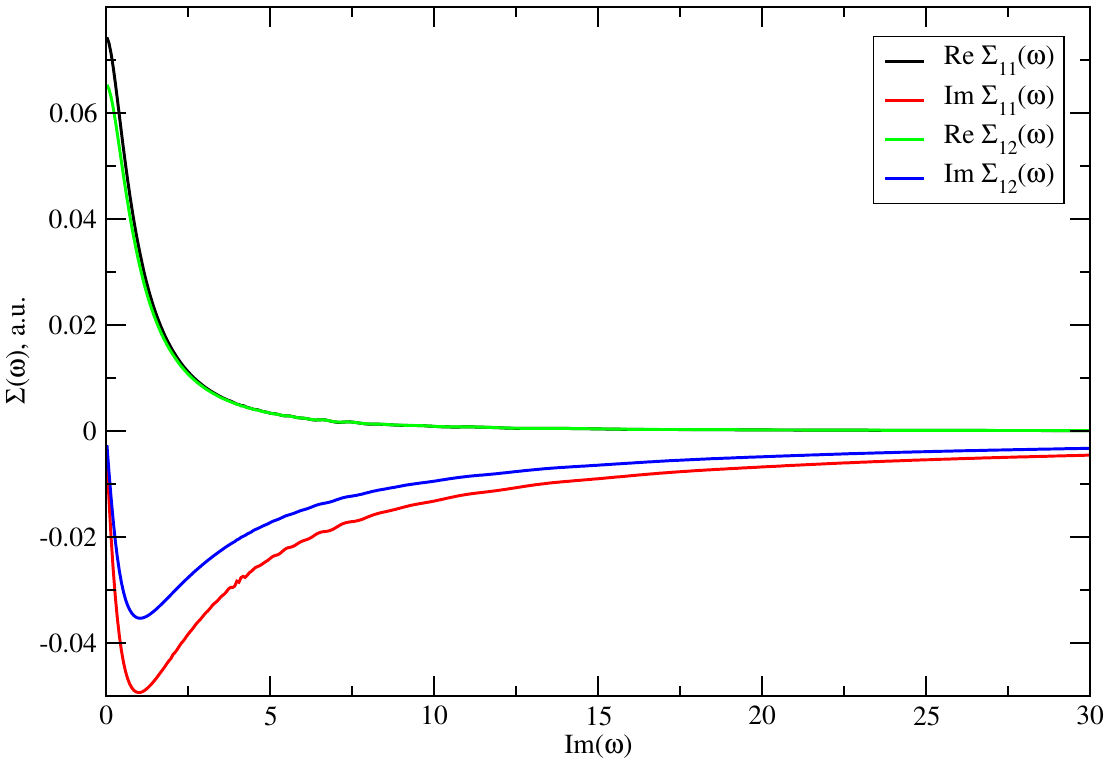}}
\caption{\label{075_2} Real-frequency spectral function and correlated bands, Matsubara self-energy for ``solution~2'' at $R=0.75$~\AA.}
\end{figure}
\begin{figure}[htp]
\centering
\subfigure{\includegraphics[scale=1.0]{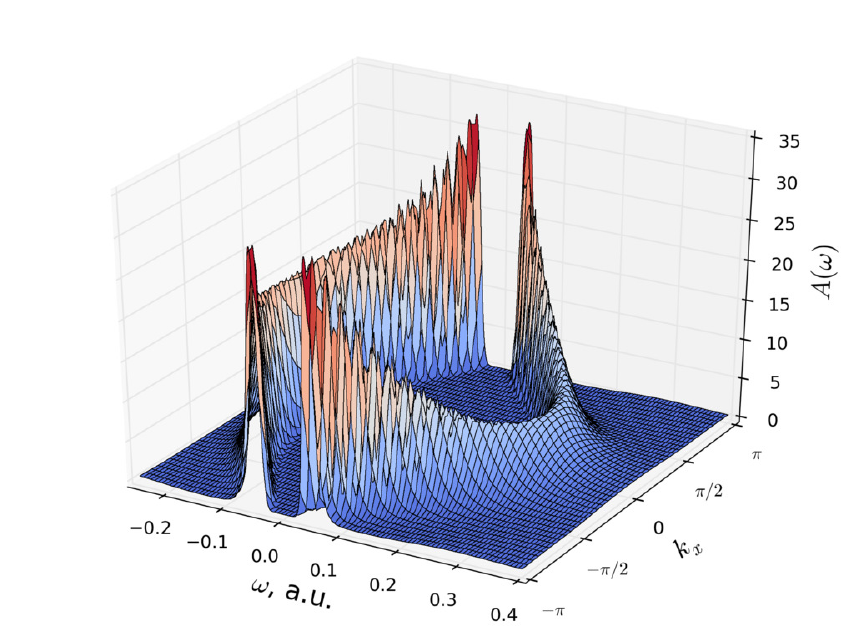}} \\
\qquad
\subfigure{\includegraphics[scale=0.9]{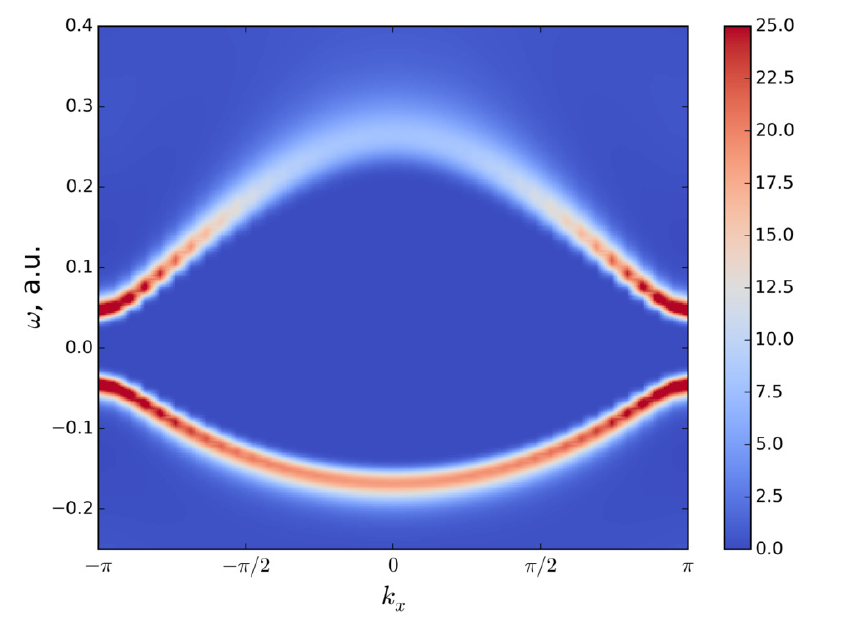}} \\
\qquad \\
\qquad \\
\subfigure{\includegraphics[scale=0.7]{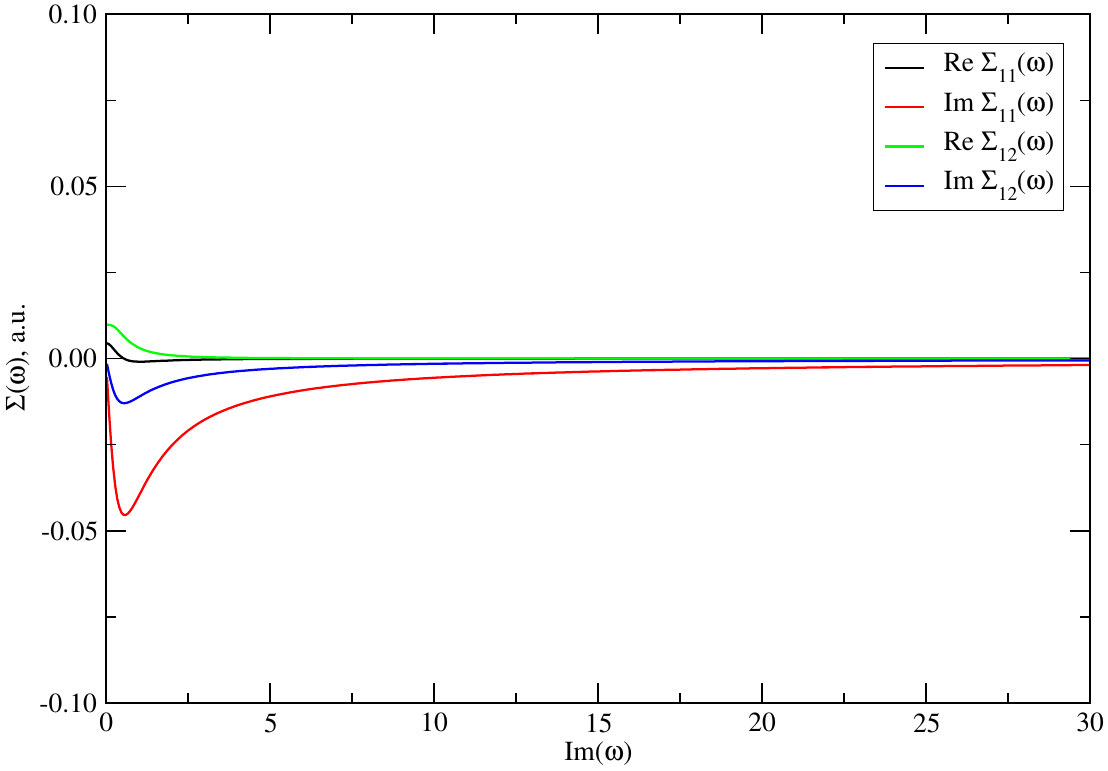}}
\caption{\label{175_1} Real-frequency spectral function and correlated bands, Matsubara self-energy for ``solution~1'' at $R=1.75$~\AA.}
\end{figure}
\begin{figure}[htp]
\centering
\subfigure{\includegraphics[scale=1.0]{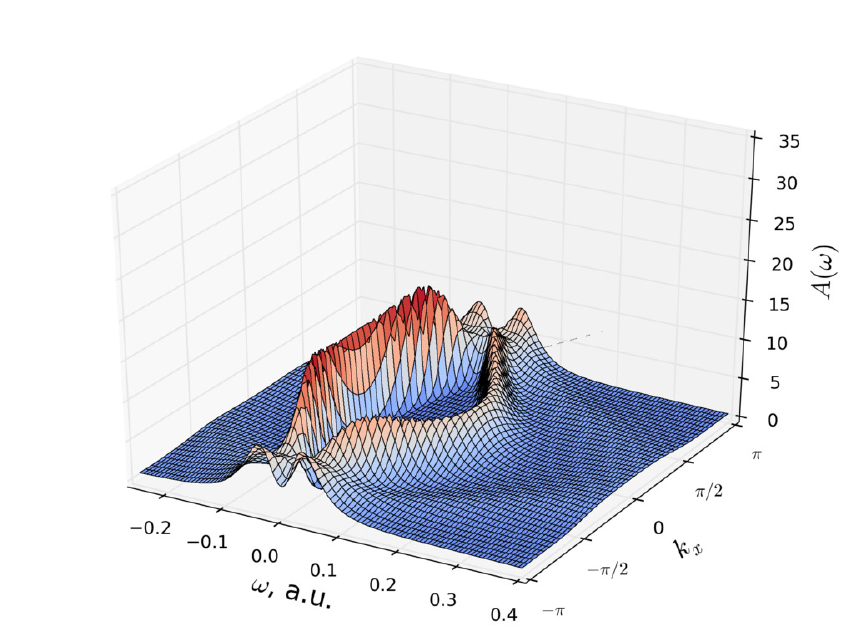}} \\
\qquad
\subfigure{\includegraphics[scale=0.9]{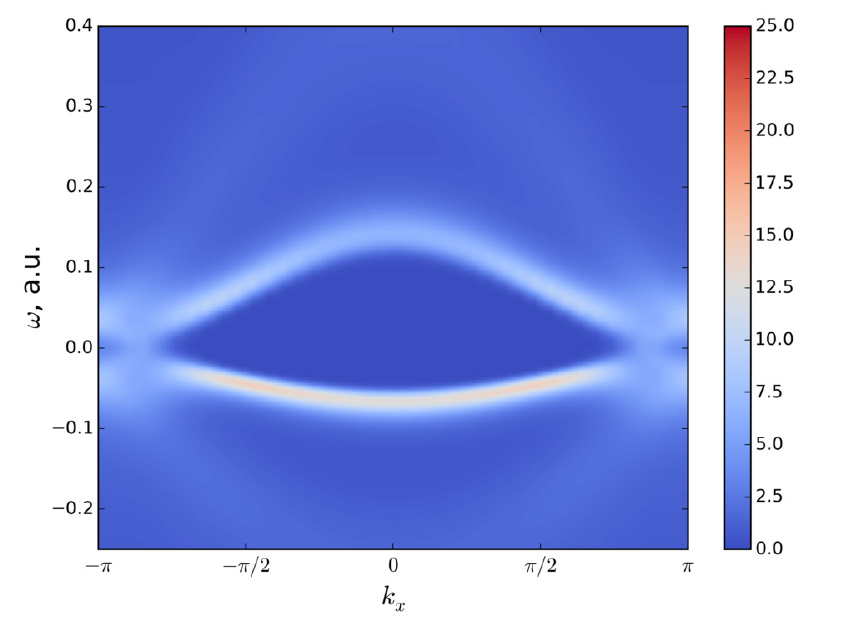}} \\
\qquad \\
\qquad \\
\subfigure{\includegraphics[scale=0.7]{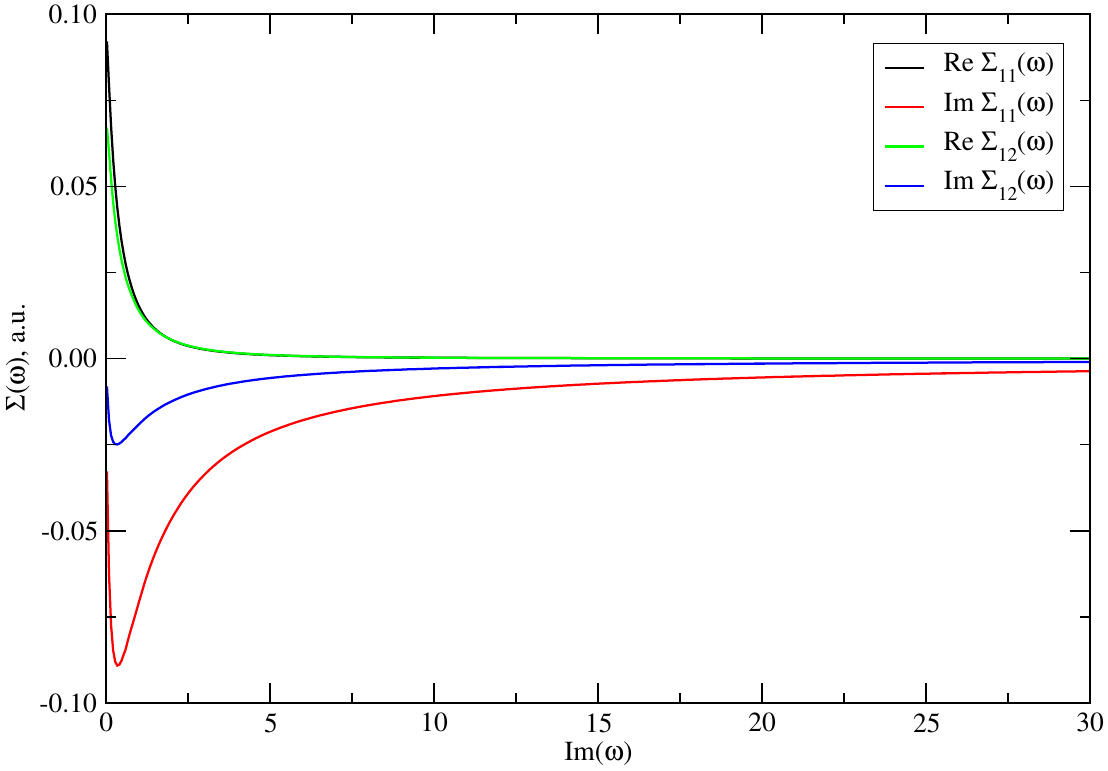}}
\caption{\label{175_2} Real-frequency spectral function and correlated bands, Matsubara self-energy for ``solution~2'' at $R=1.75$~\AA.}
\end{figure}
\begin{figure}[htp]
\centering
\subfigure{\includegraphics[scale=1.0]{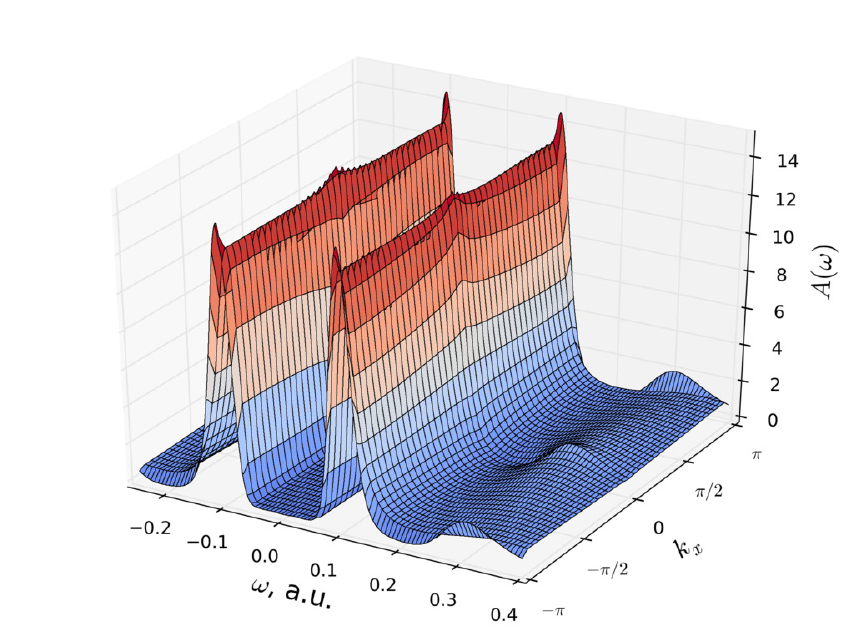}} \\
\qquad
\subfigure{\includegraphics[scale=0.9]{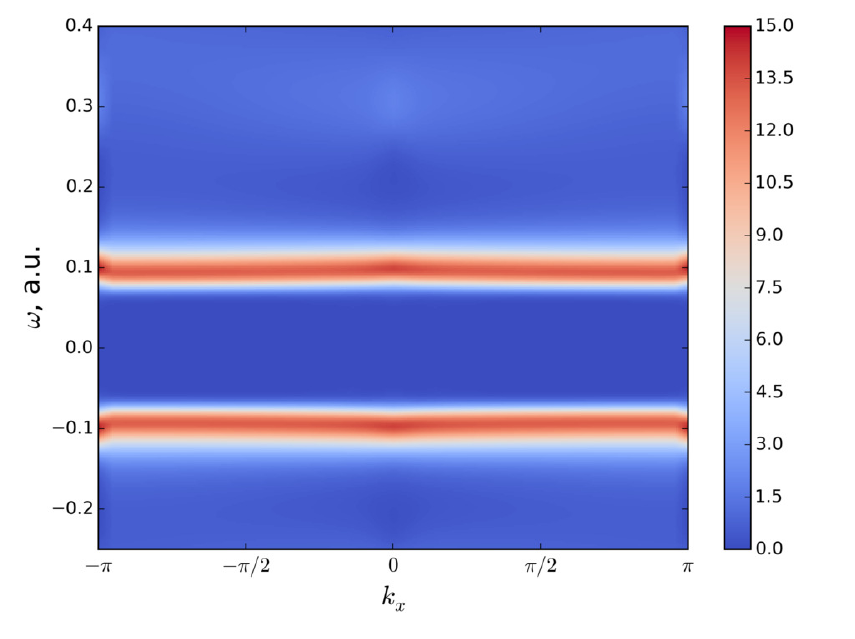}} \\
\qquad \\
\qquad \\
\subfigure{\includegraphics[scale=0.7]{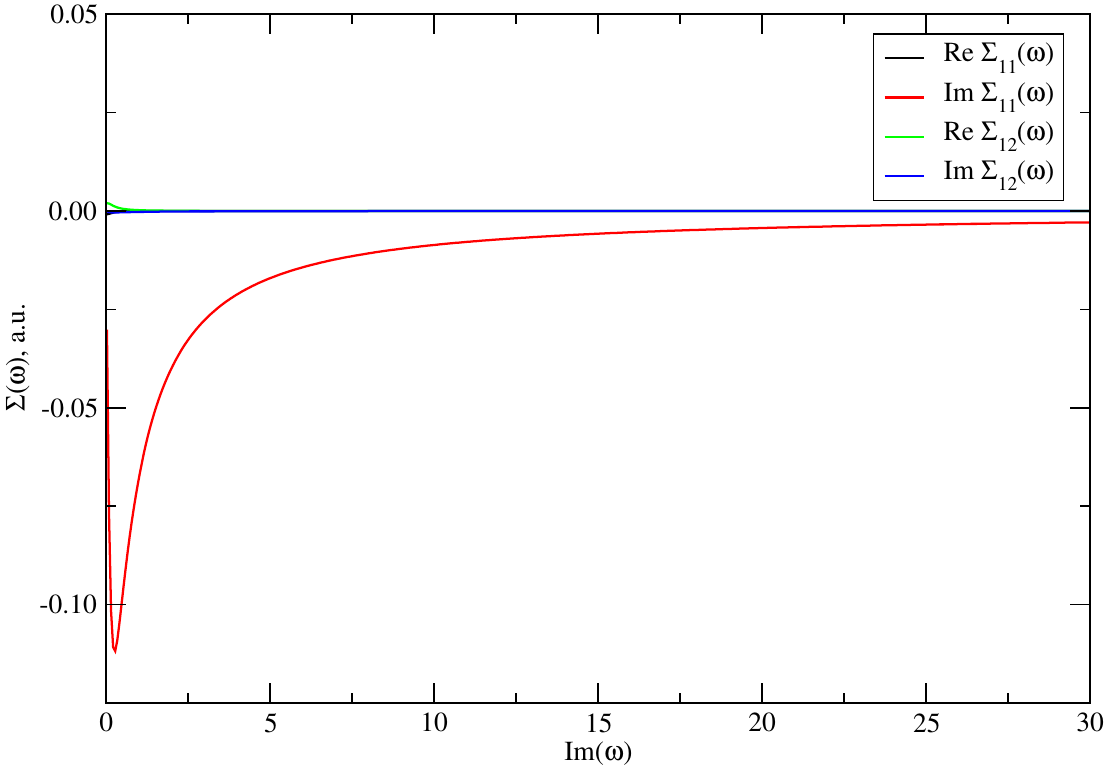}}
\caption{\label{800_1} Real-frequency spectral function and correlated bands, Matsubara self-energy for ``solution~1'' at $R=4.00$~\AA.}
\end{figure}
\begin{figure}[htp]
\centering
\subfigure{\includegraphics[scale=1.0]{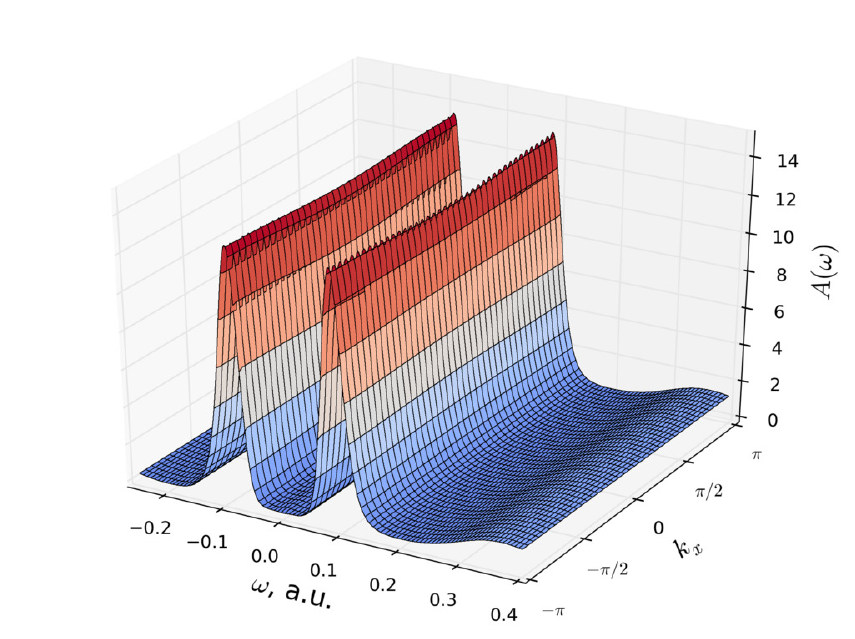}} \\
\qquad
\subfigure{\includegraphics[scale=0.9]{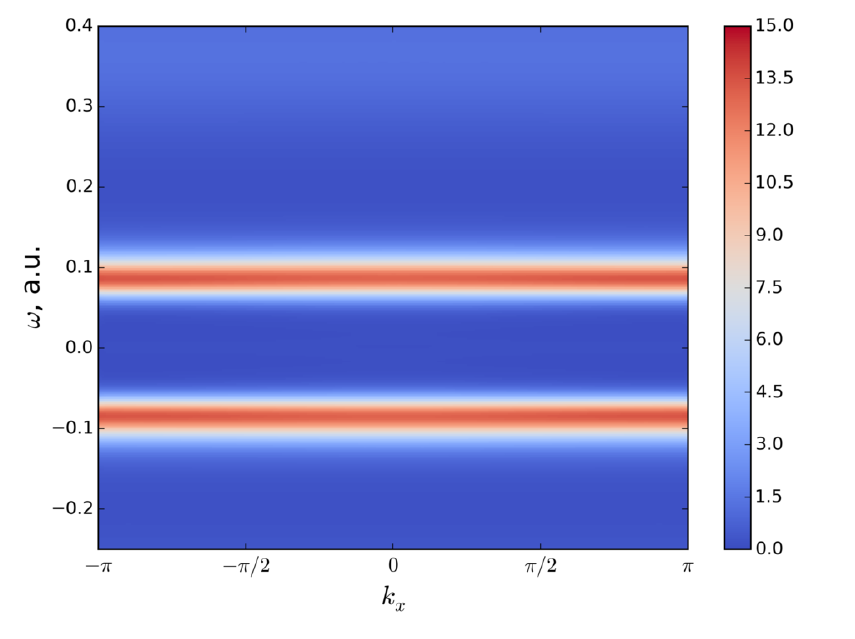}} \\
\qquad \\
\qquad \\
\subfigure{\includegraphics[scale=0.7]{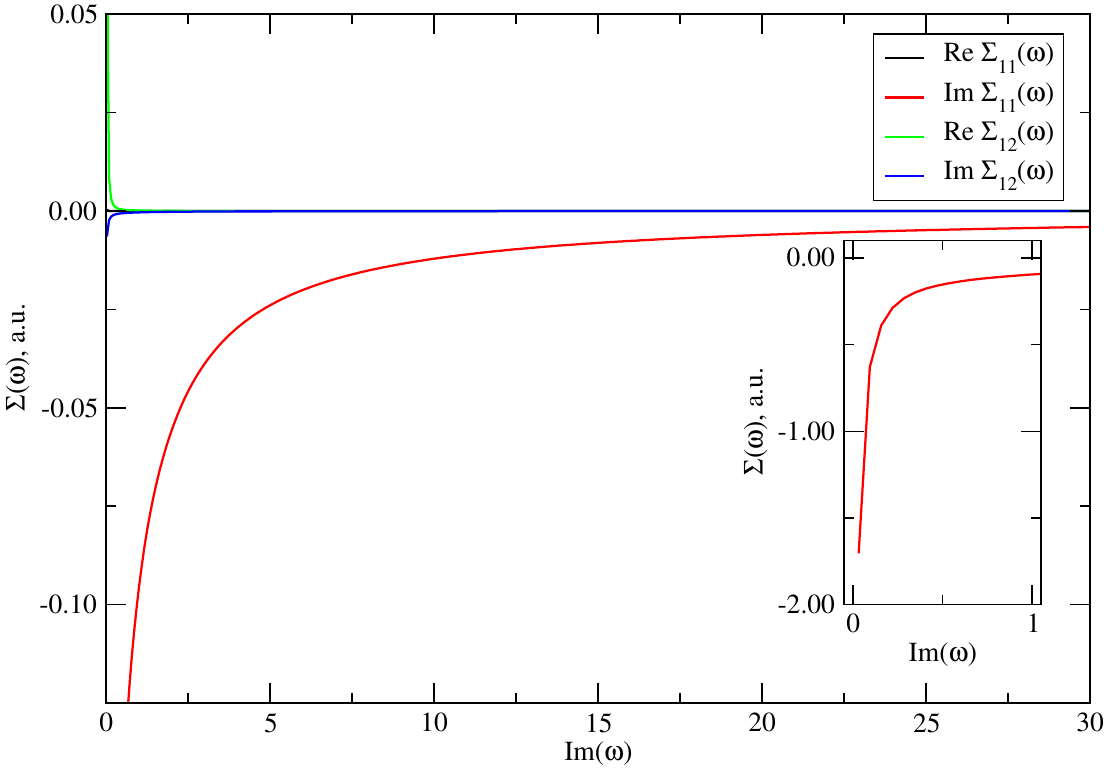}}
\caption{\label{800_2} Real-frequency spectral function and correlated bands, Matsubara self-energy for ``solution~2'' at $R=4.00$~\AA.}
\end{figure}

\subsubsection{Short bond length/high pressure}

In the condensed regime, $R=0.75$~\AA, the green curve (``solution~1'') displays metallic behavior with two peaks of 
the spectral function $A(\mathbf{k}, \omega)$ at $k=\pm\pi$ for $\omega=0$, as can be seen from Fig.~\ref{075_1}.
The resulting spectral function projection shows that the band gap closes at $k=\pm\pi$ as expected from a metallic solution. 
The Matsubara self-energies (the third panel in Fig.~\ref{075_1} contains the diagonal $\Sigma_{11}(\omega)$ and off-diagonal  $\Sigma_{12}(\omega)$ elements, where $\omega$'s are \textit{imaginary} frequencies $\omega_n=i(2n+1)\pi/\beta$) for ``solution~1'' display a typical Fermi liquid behavior that is characteristic for both metals and band insulators. 
The metallic behavior of ``solution~1'' is confirmed by the spectral function showing a non-zero density of states at the Fermi energy.

The other GF2 solution, ``solution~2'', for $R=0.75$~\AA~illustrated in Fig.~\ref{075_2}, 
also displays a metallic character and shows no gap for $k=\pm\pi$ at $\omega=0$. 
However, this spectral function looks different than the one for ``solution~1''. 
The self-energy for ``solution~2'' also has a Fermi liquid character with 
the imaginary part very similar to ``solution~1'', while small differences between the two solutions are visible in the real part.
As we have mentioned previously, an entropy contribution is necessary to determine which of the two solutions
gives rise to the most stable metallic phase at a given temperature.

\subsubsection{Intermediate bond length/intermediate pressure} 

The intermediate bond length regime, $R=1.75$--2.00~\AA, is very interesting since both solutions display qualitatively different behaviors. 
As illustrated in Fig.~\ref{175_1}, the spectral function of ``solution~1'' clearly shows an insulating behavior with two bands separated by a band gap. 
The self-energy of ``solution~1'' is not divergent near zero frequency; consequently, the phase that emerges is a band insulator and not a Mott insulator.

The spectral function of ``solution~2'' in Fig.~\ref{175_2} is gapless for $\omega=0$ indicating a metallic phase. 
The self-energy of ``solution~2'' has a Fermi liquid profile and confirms the metallic character of this solution.

\subsubsection{Long bond length/low pressure} 

In the stretched regime, $R=4.00$~\AA, both solutions correspond to insulators.
The spectral function of ``solution~1'', illustrated in Fig.~\ref{800_1}, displays an insulating behavior
and its self-energy confirms its band insulator character.

The spectral function of  ``solution~2'' in Fig.~\ref{800_2} has two sharp peaks near $\omega=-0.1$ and $0.1$ a.u. 
The self-energy of ``solution~2'' is large for the frequencies near $\omega=0$ confirming a Mott insulator behavior.
A Mott insulator phase for 1D periodic hydrogen lattice can be described by an open-shell singlet wavefunction encompassing all hydrogen atoms in the crystal.

Let us stress here that recovering the Mott behavior requires accounting for the multi-reference character of the electronic state. 
Consequently, a zero-temperature perturbation theory such as MP2 would experience divergencies 
and would not be able to illustrate the Mott phase. 
The GF2 success in qualitatively capturing this phase stems from its iterative character and 
inclusion of higher order diagrams than the ones originally included 
in the 2nd-order self-energy expansion.

\subsection{Occupation numbers in the $\mathbf{k}$-space}

Natural occupation numbers, \textit{i.e.}, the eigenvalues of the density matrix, provide an additional insight into electronic correlations and phase transitions present in the system. 
Since we work in the non-orthogonal AO basis, 
we calculate the density matrix in the L\"owdin orthogonalized basis at every $\mathbf{k}$
\begin{equation}\label{Psao}
\mathbf{\gamma}^{\mathbf{k}}_{orth} = (\mathbf{S^k})^{-1/2}\mathbf{\gamma^k}(\mathbf{S^{k\dagger}})^{-1/2}
\end{equation}
and diagonalize it to obtain natural $\mathbf{k}$-space occupation numbers.
Generally, different numbers of $\mathbf{k}$-points are required to achieve convergence at different geometries, and the number
of the density matrix eigenvalues changes with the interatomic distance.

\begin{figure}
\includegraphics[scale=0.7]{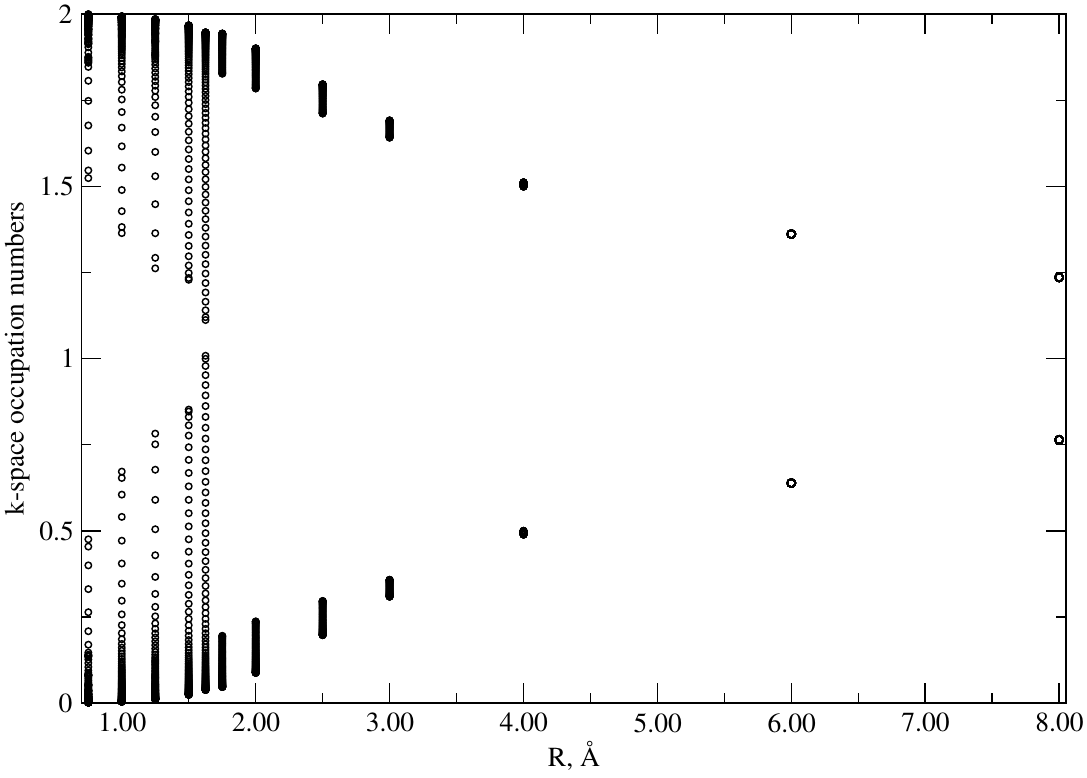}
\caption{\label{occp} Equidistant 1D hydrogen: $\mathbf{k}$-space natural occupations for ``solution~1''.}
\end{figure}

The eigenvalues of $\mathbf{\gamma}^{\mathbf{k}}_{orth}$ along the ``solution~1'' curve are plotted in Fig.~\ref{occp} for all momentum vectors $\mathbf{k}$ used to sample the first Brillouin zone. 
The behavior of these natural occupations starkly contrasts the finite system behavior analyzed in Ref.~\onlinecite{GF2_1}.
In the short bond regime (0.75--1.75~\AA)
the infinite 1D hydrogen lattice features a much wider spectrum of 
fractional occupations. 
Thus, even around equilibrium the ground state of the 1D infinite hydrogen lattice 
acquires some multi-reference character.
At $R=1.75$~\AA,~the natural occupations discontinuously change  from a wide spectrum to two relatively narrow clusters 
in the vicinity of 0 and 2 indicating a possible phase transition.
Upon further stretching, these clusters shrink and depart from 0 and 2 occupations, but never collapse to 1 that would be indicative of a Mott regime.

\begin{figure}
\includegraphics[scale=0.7]{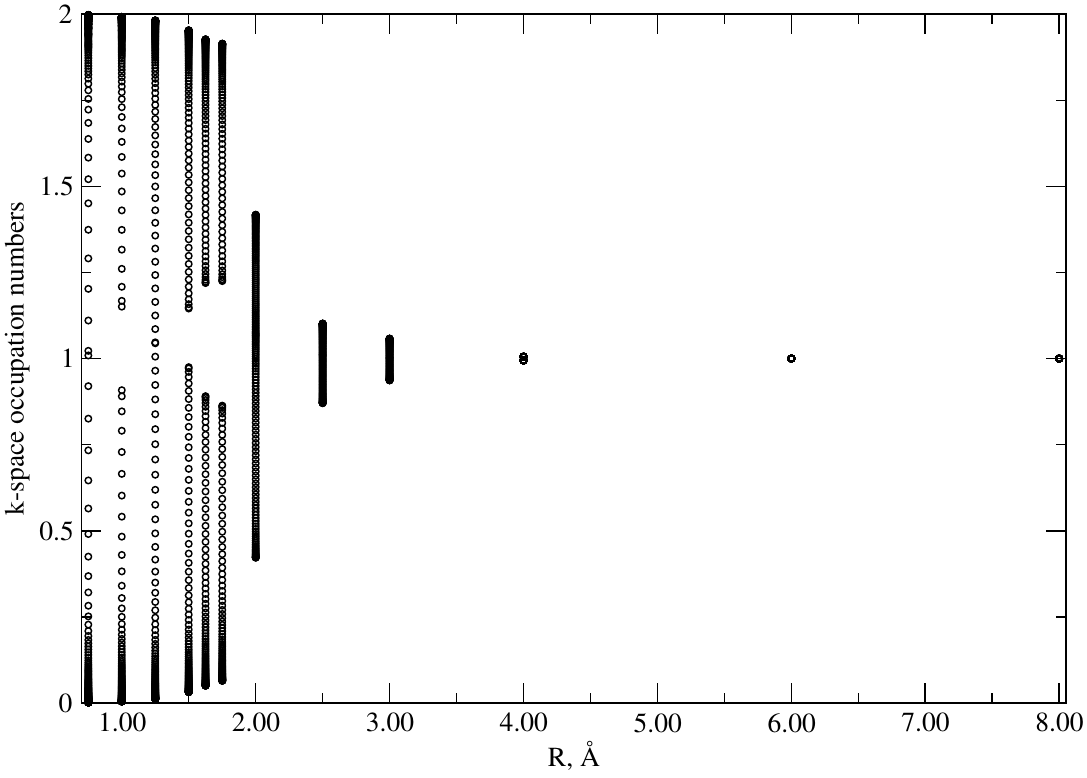}
\caption{\label{occp1} Equidistant 1D hydrogen: $\mathbf{k}$-space natural occupations for ``solution~2''.}
\end{figure}

The occupation numbers in Fig.~\ref{occp1} corresponding to ``solution~2'' (blue curve in Fig.~\ref{energy})
also display a multi-reference nature 
of the electronic state around equilibrium. In contrast to ``solution~1'', however,
for interatomic distances beyond 2~\AA~the occupation
spectrum rapidly collapses to 1 indicating a clear phase transition to the strongly correlated Mott phase.

\subsection{Peierls distortion of the 1D infinite hydrogen lattice}
\begin{figure}
\includegraphics[scale=0.7]{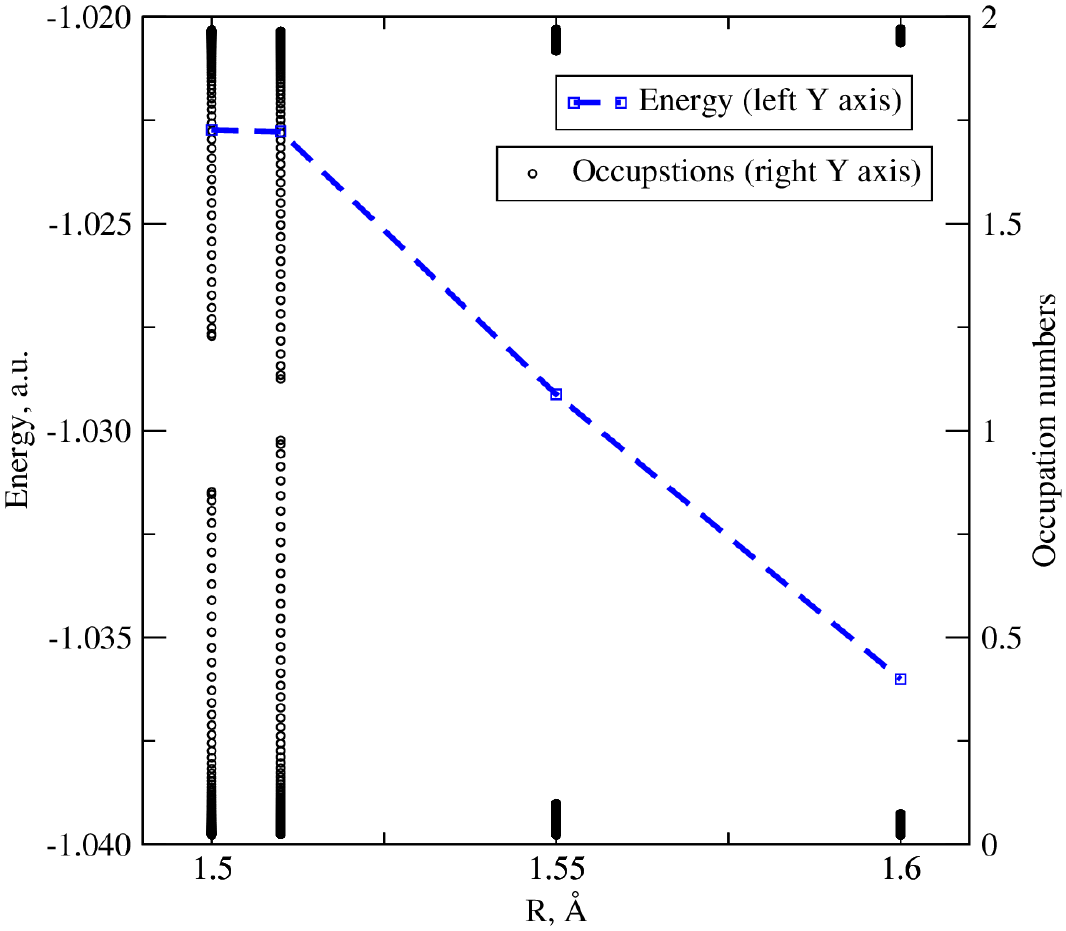}
\caption{\label{Peierls} Energies and $\mathbf{k}$-space natural occupations under the Peierls distortion of the 1.50~\AA~equidistant 1D hydrogen lattice.}
\end{figure}

Equidistant 1D hydrogen lattices are known to undergo the Peierls distortion,~\cite{PhysRevB.50.14791} \textit{i.e.}, they energetically favor an alternant pattern of bond lengths to an equidistant one. 
In Fig.~\ref{Peierls}, we plot the energies and natural occupation numbers for the 1.50~\AA~equidistant 1D hydrogen lattice undergoing a slight distortion 
resulting in an interatomic distance pattern $R+\delta R$ --- $R-\delta R$. 
A small alteration of the interatomic separation pattern by $\delta R = 0.05$~\AA~results in a discontinuous change in the natural $\mathbf{k}$-occupations, indicative of a phase transition, 
accompanying the energy decrease. 

\begin{figure}
\begin{center}
\includegraphics[scale=0.6]{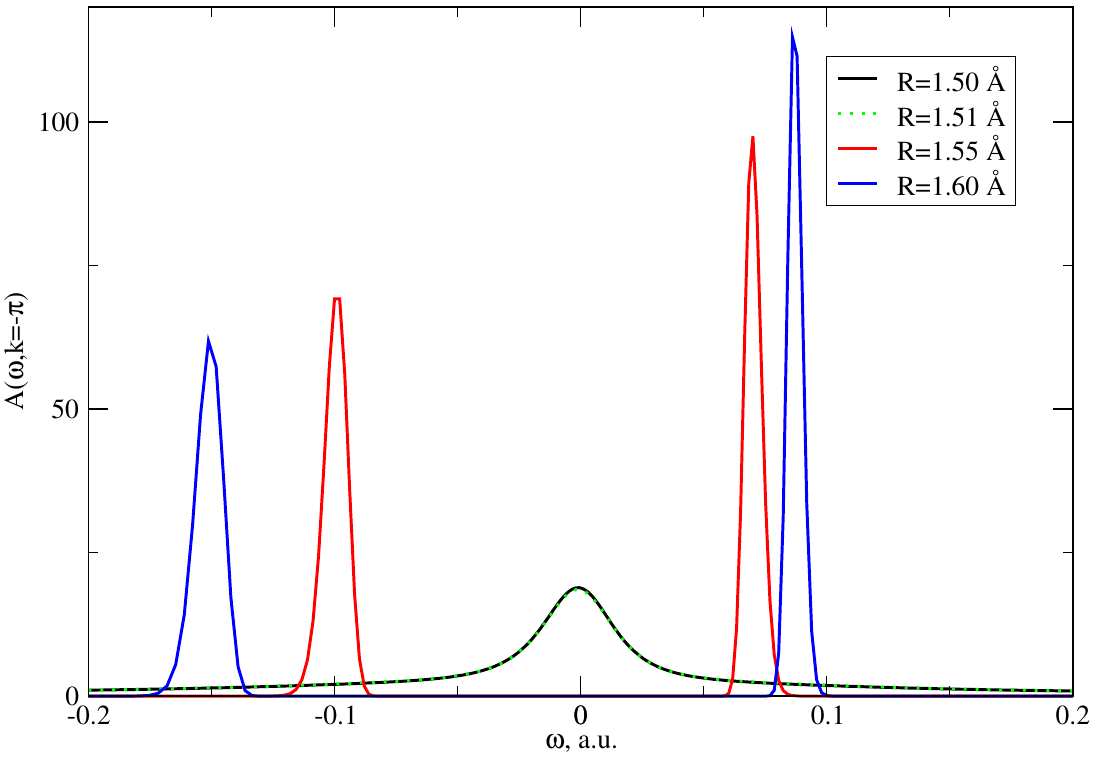}
\end{center}
\caption{\label{gap_open} Gap opening in 1D hydrogen at $R = 1.50+\delta R$~\AA.}
\end{figure}

Indeed, while the occupation numbers of equidistant 1D hydrogen lattices in Figs.~\ref{occp} and \ref{occp1} display a noticeable multi-reference character for 
distances below 2~\AA, the $\delta R = 0.05$~\AA~distorted lattice is quite weakly correlated and has occupations 0 and 2 for distances greater than
1.52~\AA.
Moreover, the appearance of the spectral functions (Fig.~\ref{gap_open}) at $k=-\pi$ for several $\delta R$
clearly indicates a metal to band insulator transition under such distortion.

\subsection{The importance of GF2 iterations}

In Table~\ref{Energy_iter}, we list the unit cell energy and density matrix elements for the $R=0.75$~\AA~equidistant lattice in the course of GF2 iterations.
\begin{table}[h!]
\centering
\begin{tabular}{c|c|c|c} 
 \hline
$iteration$ & E, a.u. & $\gamma_{11}^{00}$ & $\gamma_{12}^{00}$  \\ 
 \hline\hline
Hartree--Fock&	$-0.94477$&      0.905  &   0.358     \\ 
first &	$-0.96202$&      0.954  &   0.520    \\       
last (converged)&	$-0.96272$   &   0.954  &   0.251    \\  
 \hline
\end{tabular}
\caption{Equidistant 1D hydrogen lattice, $R = 0.75$~\AA, 2 atoms per unit cell: reference cell energy and density matrix elements in the course of GF2 iterations.}
\label{Energy_iter}
\end{table}
The essential observation is that the convergence is not reached in a single iteration, 
but it takes several GF2 cycles before the unit cell energy and density matrix elements stabilize.
\begin{figure}
\begin{center}
\includegraphics[scale=0.7]{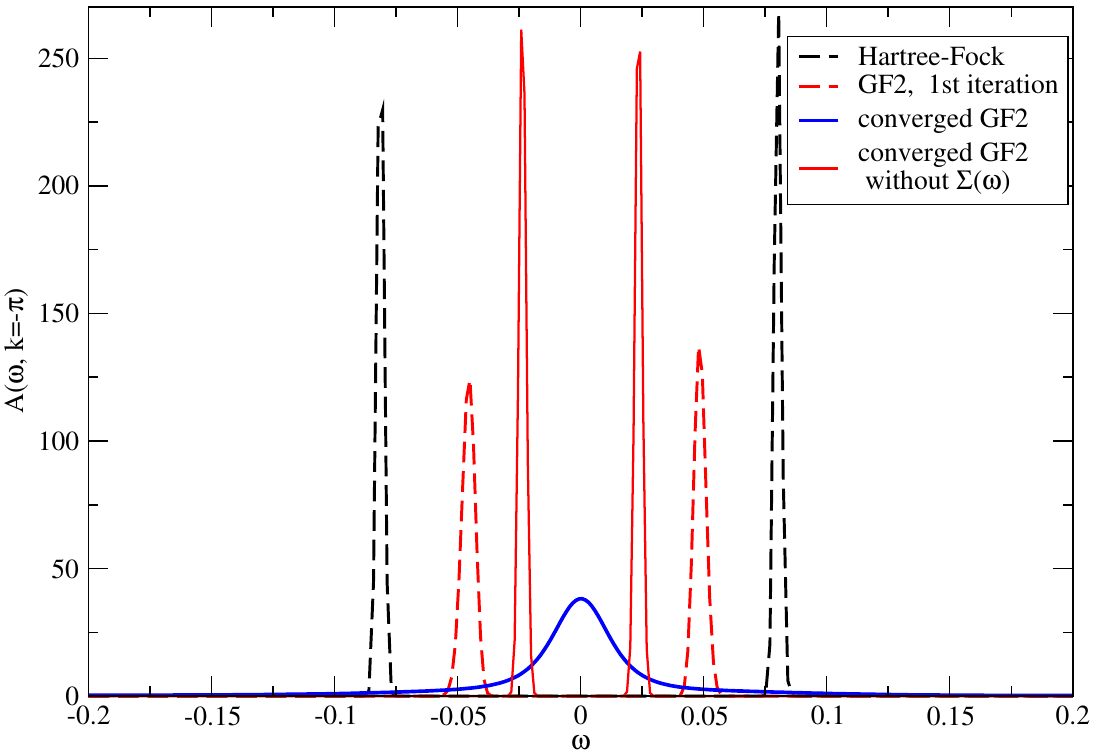}
\end{center}
\caption{\label{gap_p} $A(\omega, k=-\pi)$ at $R = 0.75$~\AA~with the inclusion of various elements of the GF2 procedure.}
\end{figure} 

To further highlight the importance of executing the GF2 cycle in a self-consistent fashion, updating both the frequency-dependent and frequency-independent 
parts of the self-energy, we consider the behavior of the spectral function of the same lattice in the course of the iterations. 
Such lattice is metallic at the convergence of iterations for both GF2 solutions. 
In Fig.~\ref{gap_p},
we show the ``slices'' of the spectral functions at $k=-\pi$ corresponding to the HF reference, single iteration of GF2, 
correlated Green's function at convergence containing the $\mathbf{\Sigma_{\infty}}$ term and neglecting the
frequency-dependent term $\mathbf{\Sigma}(\omega)$, and finally the fully converged GF2. 

The metallic behavior transpires only at the level of the fully self-consistent GF2 solution 
with electron correlations incorporated in both the updated $\mathbf{\Sigma_{\infty}}$ term and
the explicitly frequency-dependent $\mathbf{\Sigma}(\omega)$ part of the self-energy. 
Whenever the full iterative cycle is not performed, qualitatively different gapped solutions appear.

\subsection{Convergence of GF2 for metallic solutions}
We briefly comment on the convergence of the GF2 method for metallic systems.  Zero-temperature 2nd-order MP2 is known to diverge for metals with increasing number of $\bf k$-points as was theoretically demonstrated in Ref.~\onlinecite{Fetter-Walecka}, p. 32 (see also Refs.~\onlinecite{Jishi, :/content/aip/journal/jcp/140/12/10.1063/1.4867783, PhysRevLett.110.226401}) and numerically observed by Hirata~\cite{:/content/aip/journal/jcp/143/10/10.1063/1.4930024} and Kresse.~\cite{:/content/aip/journal/jcp/133/7/10.1063/1.3466765}
RPA does not encounter these divergencies~\cite{PhysRevLett.110.226401, PhysRevB.87.075111} due to the inclusion of an infinite series of different order ``bubble" diagrams.

In contrast to MP2, in GF2 we do not observe any divergence for metallic solutions of the 1D hydrogen lattice.
As shown in Tab.~\ref{metallic_conv} for $R = 0.75$~\AA, the total unit cell energy and the $E_{2b}$ term (defined in Eq.~\ref{Galitskii}) are convergent with respect to the number of $\bf k$-points.
\begin{table}[h!]
\centering
\begin{tabular}{c|c|c} 
 \hline

\# of $\bf k$-points &                        total energy   &                 2-body energy     \\ 
 \hline
  \hline

226     &                          $-0.962719$      &                  $-0.035374$   \\

400       &                        $-0.962685$         &                $-0.035818$      \\  

600          &                     $-0.962715$           &               $-0.035502$  \\     

800             &                  $-0.962715$             &                $-0.035502$  \\ 
 \hline
\end{tabular}
\caption{Unit cell total energies and 2-body electronic energies calculated using different numbers of $\bf k$-points for the metallic solution of the 1D equidistant periodic hydrogen, $R = 0.75$~\AA. Given $k$'s correspond to the number of $\mathbf{k}$-points in the inversion-unique half of the first Brillouin zone; for each $k$ there are $2k-1$ points in the full first Brillouin zone. $k=226$ is sufficient to attain (standard) \textsc{gaussian} convergence requirements.}
\label{metallic_conv}
\end{table}

Several key differences to the MP2 case exist. The  iterative nature of GF2 leads to the inclusion of a series of increasingly higher-order diagrams. We surmise that while some of these  diagrams are divergent 
in the zero momentum limit, the sum remains convergent. This infinite series is similarly convergent as the series of ``bubbles" in RPA, and its inclusion appears to cure GF2 from the MP2-type of divergencies. Note, however, that
the series included in GF2 is different than the infinite ``sum of bubbles" present in RPA.
Additionally, the HF solution of the 1D hydrogen lattice is insulating for the distances studied, and the metallic character of the solution 
is only generated due to  the GF2 self-consistency.
While the 1D case is special, MP2 divergencies for metallic solutions have been observed for 1D cases at finite temperature.~\cite{:/content/aip/journal/jcp/143/10/10.1063/1.4930024}  Consequently, in the future, it is desired to investigate theoretically or at least numerically the behavior of GF2 for the metallic solutions in 2D and 3D cases. At present, we do not have a proof that such behavior is generally convergent with the number of $\bf k$-points.

\section{Conclusions}\label{conclusions}

We have presented a prototype implementation of the finite-temperature 
iterative 2nd-order Green's function theory for periodic systems.

Our GF2 implementation is based on the evaluation of the self-energy in the real space using atomic orbitals
and solving the Dyson equation in $\mathbf{k}$-space.
Consequently, the resulting self-energy is local and decays fast with respect to the number of cells included.
This fast decay of the AO self-energy can be used to increase computational robustness of our implementation.
The evaluation of the Dyson equation in the reciprocal space via independent matrix inversions 
at every $\mathbf k$-point facilitates an efficient evaluation 
of the Green's function.

To learn about the performance of GF2 for solids, we have analyzed a 1D infinite hydrogen lattice. 
Our simulations at $T \approx 3160$K ($\beta = 100$[1/a.u.]) for each interatomic distance
converge to two phases different in internal energy, spectral properties, or both. 

We have confirmed that GF2 is capable of describing metallic phases of 1D hydrogen lattice, and it is not affected by the divergencies typical for the zero-temperature MP2.

GF2 is capable of qualitatively describing a metal-to-insulator transition. The resulting Mott phase that requires a multi-determinantal wavefunction can 
be qualitatively captured, and the 1D hydrogen lattice can be qualitatively correctly atomized. 
We attribute the emergence of the Mott phase in GF2 to the
iterative nature of the method implicitly accounting for the multi-determinantal wavefunction
via a self-consistent update of the Fock-type reference using the correlated density matrix.
This appears sufficient to qualitatively cure the known deficiencies of single-reference methods, including MP2, in stretching and breaking bonds.

Our discernment of various phases of equidistant 1D hydrogen relies on a careful examination of the behavior
of the natural occupations, self-energy, and spectral function. The $\mathbf{k}$-dependent spectral function obtained
via the analytical continuation of the Matsubara Green's function to the real axis, coupled with the self-energy behavior analysis
allows us to recognize metallic, band insulator, and Mott phases.

To the best of our knowledge, GF2 is one of the very few available \textit{ab initio} methods to account, at least qualitatively, for 
weak and strong electronic correlations simultaneously that can be applicable to metals, band and Mott insulators.
Therefore, a natural question arises if GF2 can be applied to large systems with multiple orbitals in the unit cells.
The formal scaling of periodic GF2 is $O(N_{orb}^5N_{cell}^4n_{\tau})$ where $N_{orb}$, $N_{cell}$, and $n_{\tau}$ are, respectively, the
number of orbitals in the unit cell, the number of unit cells included in the evaluation of the self-energy, and the size of the imaginary-time grid.
Though the calculation is easily parallelized over the imaginary-time points, it can be 
expected that GF2 is still more expensive than traditional MP2. 
For this reason, further development of GF2 applications to systems with large numbers of orbitals
per unit cell can benefit from employing stochastic techniques to evaluate the 2nd-order self-energy.~\cite{doi:10.1021/jz402206m, PhysRevLett.113.076402, doi:10.1021/ct300946j, :/content/aip/journal/jcp/140/3/10.1063/1.4862255, :/content/aip/journal/jcp/138/16/10.1063/1.4801862, 
:/content/aip/journal/jcp/137/20/10.1063/1.4768697}
Efforts in this direction are underway.

\section{Acknowledgements}
D.Z. and A.A.R. would like to acknowledge the DOE grant No. ER16391. A.A.R. is grateful to Prof. Emanuel Gull for invaluable help 
with analytical continuation, Dr. Jordan J. Phillips for extremely useful molecular GF2 reference data and multiple enlightening discussions, and Alexei A. Kananenka for critically reading the manuscript.

\end{document}